\documentclass[journal]{IEEEtran}

\ifCLASSINFOpdf

\else
 
\fi

\usepackage[utf8]{inputenc}
\usepackage{graphicx}
\graphicspath{{Images/}}
\usepackage{caption}
\usepackage{multirow}
\usepackage{tabu}
\usepackage{color}
\usepackage{array}
\usepackage{amsmath,mathtools}
\usepackage{enumitem}
\usepackage{siunitx}
\usepackage{dirtytalk}
\usepackage{amssymb}
\usepackage{blkarray,booktabs,bigstrut}
\usepackage{titlesec}
\usepackage{derivative}
\usepackage{amssymb}
\usepackage{booktabs,chemformula}
\usepackage{lipsum}
\usepackage{algorithm}
\usepackage{algorithmicx}
\usepackage{algpseudocode}
\usepackage{etoolbox}
\usepackage{textcomp}

\makeatletter
\patchcmd{\ALG@step}{\hskip\algorithmicindent}{}{}{}
\makeatother
\usepackage[font=small, justification=justified, singlelinecheck=false, width=\linewidth]{caption}
\DeclareCaptionLabelFormat{myformat}{\figurename~#2.}
\usepackage[usestackEOL]{stackengine}
\usepackage{siunitx}
\usepackage{subcaption}
\stackMath

\usepackage{caption}
\makeatletter
\DeclareCaptionLabelFormat{mysublabelfmt}{(\alph{sub\@captype})}
\makeatother
\usepackage[labelformat=mysublabelfmt]{subcaption}

\usepackage{caption}

\DeclareCaptionFormat{myformat1}{%
  \begin{varwidth}{\linewidth}%
    \centering
    #1 \\[1ex] % Adds a newline between the label and the caption text
    #3%
  \end{varwidth}%
}

\begin{document}
\title{Molecular Communication-Inspired Particle Collector-Transmitter (PaCoT) for Heavy Metal Removal from Human Circulatory System}
\author{Hilal Esra~Yaldiz,~\IEEEmembership{Student Member,~IEEE,}
        and~Ozgur B.~Akan,~\IEEEmembership{Fellow,~IEEE}
\thanks{The authors are with the Center for neXt-generation Communications (CXC), Department of Electrical and Electronics Engineering, Koç University, 34450 İstanbul, Turkey (e-mail: \{hyaldiz23, akan\}@ku.edu.tr).}
\thanks{Ozgur B. Akan is also with the Internet of Everything (IoE) Group, Electrical Engineering Division, Department of Engineering, University of Cambridge, Cambridge CB3 0FA, UK (e-mail: oba21@cam.ac.uk).}
\thanks{This work was supported in part by the AXA Research Fund (AXA Chair for Internet of Everything at Koç University).}
}

\maketitle

\begin{abstract}
This study proposes a novel molecular communication (MC)-inspired nanomachine, PArticle COllector-Transmitter (PaCoT), to remove toxic heavy metals from the human circulatory system. PaCoT collects these toxic metals and transmits them to release nodes, such as lymph capillaries, before they reach critical organs. The design incorporates key physical parameters and operates through particle reception and release mechanisms. In the reception process, described as ligand-receptor binding reactions, modeled as a continuous-time Markov process (CTMP), PaCoT uses metallothionein proteins as receptors and heavy metals (e.g., Zn, Pb, Cd) as ligands. We assume that the toxicity condition (toxic (bit-1), non-toxic (bit-0)) is encoded into the concentration of heavy metal molecules. Thus, we consider that heavy metal concentration within the MC channel (e.g., human circulatory system) employs binary concentration shift keying (binary CSK). The concentration ratio of specific heavy metals is estimated to infer toxicity, i.e., a high ratio indicates toxicity and a low ratio suggests non-toxicity. Toxicity detection is achieved by monitoring the receptor bound duration in the presence of interferers and various types of heavy metals. After detecting and collecting toxic heavy metals, PaCoT securely retains them in a liquid medium (e.g., water) until release, employing two mechanisms: (1) a single-disc viscous micropump to regulate flow rate, and (2) Brownian motion to facilitate diffusion. PaCoT's performance is evaluated through MATLAB simulations, focusing on bit error probability (BEP) of the toxicity detection method, release time of molecules from PaCoT and energy consumption.
\end{abstract}


\IEEEpeerreviewmaketitle

\section{Introduction}
\IEEEPARstart{H}{eavy} metal pollution, caused by natural events (e.g., rain) and anthropogenic activities like fertilizer use, poses serious threats to public health and environmental sustainability worlwide \cite{prakash2023nano, ekrami2022nanotechnology, baby2022nanomaterials, world2009children}. The non-biodegradable nature of heavy metals leads to accumulation in air, water, and soil, promoting bioaccumulation in the food chain and human exposure through inhalation, ingestion, or skin contact, resulting in both acute and chronic health issues \cite{ soliman2022trophic, inobeme2023recent, munir2021heavy}.

Traditional methods for heavy metal particle removal, such as adsorption and phytoremediation, often struggle with incomplete removal and potential secondary pollution \cite{gaur2021sustainable, newmanphytoremediation}. In contrast, Molecular Communication (MC)-enabled nanomachines provide a promising alternative \cite{khan2019nanoparticles, sanchez2011controlled}, through nanotechnology, as they are designed for molecular-scale tasks like sensing, processing, actuation, and communication \cite{kuscu2021internet, suda2005exploratory, akyildiz2008nanonetworks}, making them particularly well-suited for precise intrabody applications such as the detection and removal of toxic heavy metals from the circulatory system.

While conceptual studies focus on environmental remediation \cite{vilela2016graphene, zhang2019micro, maric2018nanorobots}, a noticeable research gap remains in directly reducing toxic heavy metal particles within the human body. This research introduces the PArticle COllector-Transmitter (PaCoT) to detect, collect, transport and transmit toxic heavy metal particles to disposal nodes, facilitating their removal from the bloodstream before they reach vital organs. 

This study presents a novel method for efficiently removing toxic heavy metals from the bloodstream, focusing on designing and optimizing the PaCoT system with advanced biologically-inspired reception and release mechanisms. First we establish a mathematical model to assess the PaCoT’s particle collection capacity based on molecule size and physical properties, ensuring compatibility with biological environments and MC principles. Then we model, particle reception through ligand-receptor binding reactions, where heavy metal particles and interferer molecules serve as ligands and metallothionein proteins act as receptors. This allows the PaCoT to opportunistically capture and collect heavy metals. We assume that heavy metal concentrations behave as a binary concentration shift keying (CSK) modulated signal in the MC channel (e.g., blood vessel), allowing the toxicity condition to be inferred by estimating the concentration ratio of specific heavy metals during a sampling period, with a high ratio indicating a toxic condition. Once captured, the heavy metals are securely stored within the liquid-filled PaCoT. Furthermore, we model the release process via advection-diffusion dynamics, utilizing a single-disc viscous micropump for rapid release and Brownian motion for modelling diffusion. PaCoT’s performance is evaluated via simulations, focusing on bit error probability (BEP), particle release time, and energy consumption to optimize its design for mitigating heavy metal exposure. BEP measures detection accuracy by comparing the true toxic state (high concentrations, bit-1) with the detected state (high estimated concentration ratios, bit-1).  

The remainder of the paper is organized as follows. Section \ref{section2} outlines the PaCoT design, Section \ref{section3} presents the mathematical model, Section \ref{section4} discusses the performance evaluation, and Section \ref{section5} provides the conclusion of the study.

\begin{figure*}[htp]
    \centering  
    \includegraphics[scale=1.35]{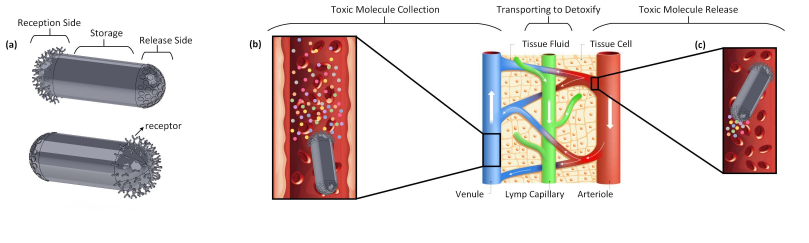}    \captionsetup{justification=centering, margin=0.1cm}    
    \caption {.\hspace{0.1 cm}Overview of PaCoT: (a) PaCoT design (b) reception and (c) release mechanisms of PaCoT.}
    \label{fig:concept}    
\end{figure*}

\section{Overview of PaCoT Mechanism} \label{section2}
The proposed idea is to create the PaCoT capable of detecting and collecting toxic heavy metals from blood vessels, then transporting and transmitting them to lymph capillaries for eventual delivery to detoxifying organs like liver and kidneys. After releasing the toxic metals, the PaCoT reenters the human circulatory system to continue detecting toxins, demonstrating its reusability. This study assumes constant blood flow, uniform pressure, and stable temperature and viscosity. The system overview is shown in Fig. \ref{fig:concept} and explained below.

\subsection{PaCoT Design Parameters for Heavy Metal Removal}
Fig. \ref{fig:concept}(a) illustrates the PaCoT design, comprising a reception side as the receiver, a storage and release side as the transmitter. The key design parameters are detailed below.
\begin{itemize}
    \item \textit{Toxic Molecule:} While certain heavy metals are essential in low concentrations, excessive exposure can lead to severe conditions like cellular dysfunction. PaCoT detects the toxicity and removes toxic heavy metals (e.g., Pb, Hg), supporting homeostasis by transporting them to release nodes for final transfer to detoxifying organs.
    \item \textit{Receptor:} Metallothioneins (MTs), cysteine-rich proteins, are crucial for metal homeostasis, detoxification, and protection against metal-induced oxidative stress through heavy metal binding. Thus, in this study, MTs are chosen as the single type of receptor for the PaCoT.
    \item \textit{Transport Protein:} Transport proteins embedded in the microchannels connecting the receptor region to the storage area capture particles released from metallothionein receptors via conformational changes.
    \item \textit{Size and Structure:} The PaCoT, designed as a rod-shaped structure with an aspect ratio of two, integrates mechanisms for reception, transmission, storage, and controlled release, considering the radii of heavy metals modeled as spherical particles based on their atomic radii and molecular masses \cite{vu2018mechanisms, jadaa2023heavy}, as detailed in Table \ref{table1}.
    \begin{table}[h]
    \centering
    \captionsetup{
        format=myformat1, 
        labelsep=none, % No separator between label and caption
        justification=centering, % Ensure caption text is centered
    }
    \caption{\textsc{Atomic Radius and Molecular Mass of Heavy Metals}}
    \label{table1}
    \begin{tabular}{c|c|c}    
      & \textbf{Atomic Radius} & \textbf{Molecular Mass} \\ \textbf{Heavy Metals} & (\textbf{\si{\pico\metre}}) &(\textbf{\si{\gram\per\mol}}) \\
    \hline
    Zinc (Zn)      & $142$ & $65.39$ \\
    \hline
    Copper (Cu)    & $145$ & $63.55$\\
    \hline
    Cadmium (Cd)   & $161$ & $112.41$\\
    \hline
    Mercury (Hg)   & $171$ & $200.59$\\
    \hline
    Lead (Pb)      & $154$ & $207.20$\\    
    \hline
    \end{tabular}
    \end{table}
    \item \textit{Material:} Polymer/graphene composites merge graphene's superior mechanical, electrical, and thermal properties with polymers' flexibility, processability, and durability. Graphene dispersion reduces aggregation and cost, while enhancing biocompatibility, making these composites ideal for biomedical applications. This study selects the polymer/graphene composite for the PaCoT material due to its synergistic performance.
    \item \textit{Mobility and Control:} We assume that self-propelled PaCoT autonomously collects toxic molecules and transports them to disposal sites, such as lymph capillaries. It uses blood flow for movement, activating propulsion only to navigate obstacles, and returns to the bloodstream after delivery to resume toxin detection and collection.
    \item \textit{Energy Harvesting:} Intrabody energy sources offer valuable harvesting potential, enabling PaCoT to generate energy from sources like blood flow and body temperature.
    \item \textit{Single-Disc Viscous Micropump:} This study examines a single-disc viscous micropump, shown in Fig. \ref{fig:SVDPump}, within a PaCoT for toxic molecule removal, providing a steady flow rate. These rotary devices use viscous forces at small scales to generate pressure, enabling precise flow control. The micropump's performance depends on geometry, rotational speed, and fluid properties. It consists of a spinning disc above a C-shaped channel, where viscous interactions drive fluid from inlet to outlet. Balancing shear-driven (Couette) and pressure-driven (Poiseuille) flows is crucial for optimizing flow rate and pressure.
    \begin{figure}[htp]
    \centering    
    \includegraphics[scale=1.25]{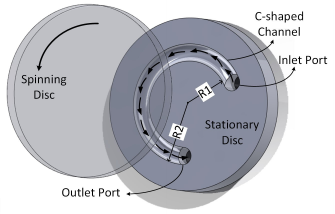}
    \captionsetup{labelformat=myformat}
    \caption{\hspace{0.1 cm}Single-disc viscous micropump.}
    \label{fig:SVDPump}   
    \end{figure} 
\end{itemize}
\subsection{Reception Process of the PaCoT}
Fig. \ref{fig:concept}(b) illustrates the PaCoT reception process. In this concept, the bloodstream, modeled as an MC channel, contains varying concentrations of $N$ types of heavy metals, which function as a binary CSK modulated signal encoding toxicity information. In the true toxicity condition, high concentrations of these metals represent a toxic condition (bit-1), while low concentrations correspond to a non-toxic condition (bit-0). These heavy metal concentrations are treated as the transmitted signal without a specific transmitter source. The PaCoT's reception side acts as a receiver, where receptors bind to heavy metal molecules and estimate the concentration ratio of each metal type (i.e., the ratio of a particular heavy metal’s concentration to the total molecular concentration near the reception area) based on receptor binding intervals. In the estimated toxicity condition, high concentration ratio of heavy metals signify toxic condition (bit-1), while lower ratios indicate a safe (non-toxic) condition (bit-0). Finally, the toxicity condition of the MC channel during the sampling period is determined by applying the decision rule to compare the true and estimated toxicity conditions.

\subsection{Release Process of the PaCoT}
After collecting toxic heavy metal particles, PaCoT stores them as suspended solids in its water-filled storage region until they reach the release nodes. Fig. \ref{fig:concept}(c) illustrates PaCoT’s release process. In this concept, the release side of PaCoT functions as a transmitter, where the captured toxic heavy metal molecules, initially gathered at the storage side, move through PaCoT’s internal structure and are ultimately released at the transmission side as transmitter molecules. During the transmission to disposal nodes, advection-diffusion dynamics come into play. Particle movement through diffusion is modeled by Brownian motion, while a single-disc viscous micropump, influenced by rotational speed, aspect ratio, and radius ratio, allows precise control over the release process.
\section{Mathematical Model of PaCoT} \label{section3}
\subsection{PaCoT Molecule Capacity Modelling}
\begin{figure}[htp]
    \centering   
    \includegraphics[scale=1.4]{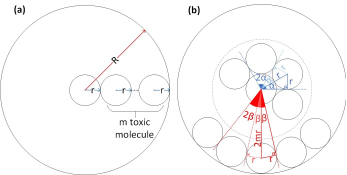}    \captionsetup{labelformat=myformat}
    \caption{\hspace{0.1 cm}The PaCoT molecular capacity. (a) Linear alignment of $m$ toxic molecules along the PaCoT radius, $R$. (b) Radial arrangement of particles around PaCoT’s center.}
    \label{fig:capacity_2D}  
\end{figure}   
The molecular capacity of PaCoT is determined by both the size of the heavy metal particles and the dimensions of PaCoT. To estimate this capacity, we assume that the molecules are arranged in a uniform pattern along the PaCoT’s cross-section, with their average radius, $r$, defined as $r = \sum_{i=1}^n x_i a_i$, where $n$ is the number of molecule types, $x_i$ represents the concentration of the $i^\text{th}$ molecule type, and $a_i$ is its radius, as shown in Fig. \ref{fig:capacity_2D}(a) and Fig. \ref{fig:capacity_2D}(b). In this configuration, the distance between the center of each particle and the PaCoT defines the hypotenuse of the triangular arrangement. 
The PaCoT’s capacity can be computed as
\begin{equation}
    C_{TM}=\frac{h}{2r}\left(1+\sum_{i=1}^m \frac{\pi}{\sin^{-1}{\frac{1}{2i}}}\right),
\end{equation}
where $m = \frac{R - r}{2r}$ represents the maximum number of molecules that can align along the PaCoT's radius, $h$ denotes its length, and $C_{TM}$ signifies the total molecular capacity.

\begin{figure}[htp]
    \centering    
    \includegraphics[scale=0.5]{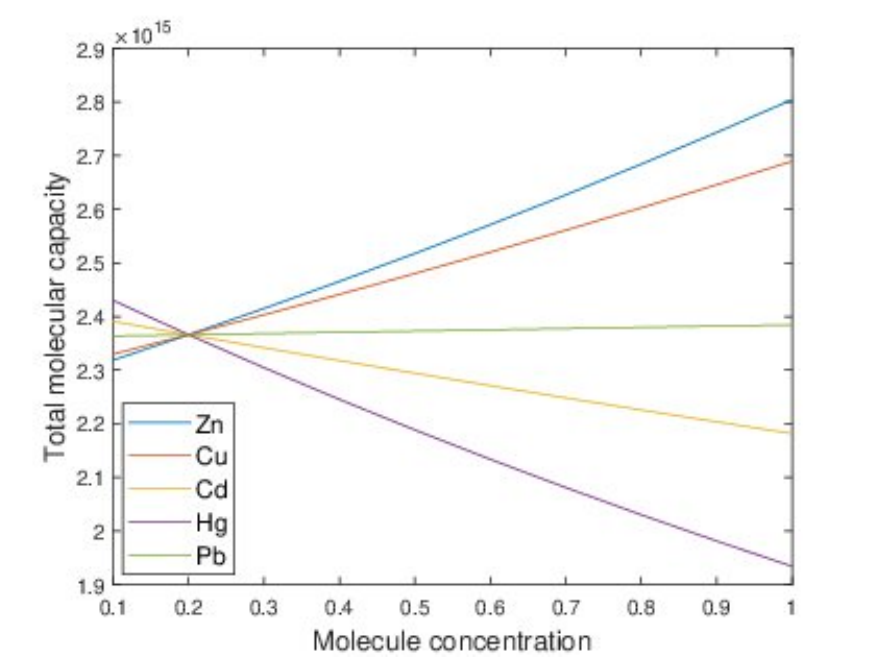}
    \captionsetup{labelformat=myformat}
    \caption{\hspace{0.1 cm}Total molecular capacity of PaCoT vs. the concentration changes of each molecule type.}
    \label{fig:capacity}   
\end{figure}

As illustrated in Fig. \ref{fig:capacity}, the PaCoT’s molecular capacity varies with the environmental concentration of different heavy metals (i.e., higher concentrations of larger particles, such as cadmium (Cd) and mercury (Hg), reduce its ability to accommodate additional toxic molecules).
\subsection{Mathematical Model of PaCoT Reception Mechanism}
Accurate estimation of heavy metal concentration ratios is crucial for determining the toxicity condition in the MC channel at each sampling instance. This section presents a mathematical model for this estimation and establishes a decision rule for identifying toxicity conditions. 

Ligand-receptor binding reactions describe how receptors interact with ligands based on binding affinity and concentration \cite{kuscu2022detection}. These interactions are modeled by the bound ($B$) and unbound ($U$) states of receptors \cite{berg1977physics, ten2016fundamental}, represented as a Markov process, which captures the random interactions between heavy metal particles and metallothionein proteins. The memoryless nature of Markov processes leads to exponentially distributed dwell times in these states \cite{liggett2010continuous}. The transitions between bound and unbound states are governed by kinetic rate constants, as expressed in the reaction equation \cite{kuscu2022detection}
\begin{equation}
    U\xrightleftharpoons [k^-] {c_L(t)k^+} B,
\end{equation}
where $c_L(t)$ denotes the fluctuating concentration of ligands over time, $k^+$ and $k^-$ are the rates at which the ligand-receptor pair binds and unbinds, respectively.

\begin{figure}[htp]
    \centering    
    \includegraphics[scale=0.80]{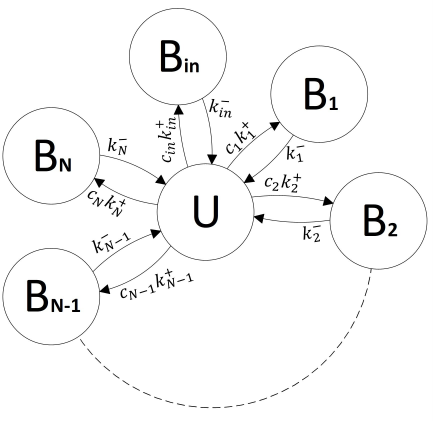}
    \captionsetup{labelformat=myformat}
    \caption{\hspace{0.1 cm}CTMP state diagram for receptor binding mechanism.}
    \label{fig:Markov}    
\end{figure}
In the presence of heavy metals (e.g., zinc (Zn)) each acting as an information molecule with distinct concentrations and binding affinities, metallothionein proteins serve as receptors that bind to these metals. The binding process with $N$ heavy metal types and interferer molecules is modeled as an $(N+2)$-state continuous-time Markov process (CTMP). The states include the unbound state $(U)$, the bound state for interferer molecules $(B_{in})$, and the bound states for each specific heavy metal $(B_1$, $B_2$, $\dots$, $B_N)$. The corresponding state diagram is shown in Fig. \ref{fig:Markov}, with the rate matrix formulated as
\begin{equation}
\small
    \boldsymbol{R} = 
    \left[\begin{smallmatrix}
        -\left( k^+_{in}c_{in} + \sum_{i=1}^N k^+_ic_i \right) & k^+_{in}c_{in} & k^+_1c_1 & \cdots & k^+_Nc_N \\
        k^-_{in} & -k^-_{in} & 0 & \cdots & 0 \\
        -k^-_1 & 0 & -k^-_1 & \cdots & 0 \\
        \vdots & \vdots & \vdots & \ddots & \vdots \\
        k^-_N & 0 & 0 & \cdots & -k^-_N
    \end{smallmatrix}\right],
\end{equation} \\
where $c_i$ denotes the concentration of each heavy metal types, while $k^+_i$ and $k^-_i$ represent the binding and unbinding rates of the $i^\text{th}$ heavy metal, respectively. The steady-state probabilities of the receptor states, denoted as $\boldsymbol{\theta} = \begin{bmatrix} p_U & p_{B_{in}} & p_1 & \dots & p_N \end{bmatrix}$, are determined by solving the linear equations $\boldsymbol{\theta} \boldsymbol{R} = 0$ and $\boldsymbol{\theta} \boldsymbol{e} = 1$, where $\boldsymbol{e}$ is a vector of ones. These equations ensure equilibrium by balancing the probabilities, with the total sum equaling one. At the equilibrium, unbound state probability is expressed as
\begin{equation}
\small
    p_U = \frac{\displaystyle k_{in}^- \prod_{i=1}^{N} k^-_i}{\displaystyle \left( k^+_{in}c_{in}\prod_{j=1}^{N} k^-_j  + \dots + k^-_{in}k^+_N c_N \prod_{j=1}^{N-1} k^-_j \right) + k^-_{in}\prod_{j=1}^{N} k^-_j}.
\end{equation}
At equilibrium, the probability of a receptor being in the bound state represents the likelihood of metallothionein proteins binding to heavy metal particles. It is determined by $p_B = p_{B_{in}} + p_{1} + \dots + p_N$, and is expressed as
\begin{equation}
\label{eq:pB}
\small
\centering
    \begin{split}      
        &p_B=1-p_U \\%&=\frac{\displaystyle k^+_{in}c_{in}\prod_{j=1}^{N} k^-_j + k^-_{in}k^+_1c_1\prod_{j=2}^N k^-_j + \dots + k^-_{in}k^+_N c_N \prod_{j=1}^{N-1} k^-_j}{\displaystyle \left( k^+_{in}c_{in}\prod_{j=1}^{N} k^-_j + \dots + k^-_{in}k^+_N c_N \prod_{j=1}^{N-1} k^-_j \right) + k^-_{in}\prod_{j=1}^{N} k^-_j}\\
        & = \frac{\displaystyle k_{in}c_{in}\prod_{j=1}^N k^-_j+k^-_{in}\sum_{i=1}^N k_i^+c_i\prod_{j=i}^{i+N-2} k_{(j \bmod N)+1}^-}{\displaystyle k_{in}c_{in}\prod_{j=1}^N k^-_j+k^-_{in}\sum_{i=1}^N k_i^+c_i\prod_{j=i}^{i+N-2} k_{(j \bmod N)+1}^- + k^-_{in}\prod_{j=1}^N k^-_j}.
    \end{split}
\end{equation}
By substituting $K_D = k^-/k^+$, known as the dissociation constant, (\ref{eq:pB}) becomes 
\begin{equation}
    p_B=1-p_U=\frac{\displaystyle c_{in}/K_D^{in}+\sum_{i=1}^N c_i/K^i_D}{\displaystyle 1+c_{in}/K_D^{in}+\sum_{i=1}^N c_i/K^i_D}.
\end{equation}
$K_D$ reflects the binding strength between heavy metal particles and metallothionein proteins. In other words, a lower $K_D$ indicates a higher affinity, increasing the likelihood of the receptor being in the bound state. This relationship captures the equilibrium dynamics of ligand-receptor interactions.

The duration in which metallothionein receptors remain in bound or unbound states provides valuable information on concentration and heavy metal types. The probability of observing a sequence of $M$ independent binding and unbinding intervals at equilibrium is calculated as
\begin{equation} 
    \begin{split}
        \label{eq:1}
        p(\{\tau_b,\tau_u&\}_M) =\frac{1}{Z}  \displaystyle \exp\left({\displaystyle-\sum_{i=1}^M \tau_{u,i}\left(\sum_{j=1}^N k^+_j c_j+k_{in}^+c_{in}\right)}\right) \\
         & \times \prod_{i=1}^M \left(\sum_{n=1}^N k^+_n c_n k^-_n e^{-k^-_n\tau_{b,i}} + k^+_{in} c_{in} k^-_{in} e^{-k^-_{in}\tau_{b,i}} \right),
    \end{split}
\end{equation}
where $Z$ is the normalization factor, $\tau_{u,i}$ and $\tau_{b,i}$ denote the observed unbound and bound time durations of heavy metals, respectively \cite{kuscu2022detection}. Assuming identical binding rates and diffusion coefficients simplifies the likelihood function as
\begin{equation}
    p(\{\tau_b,\tau_u\}_M)=\frac{1}{Z} \displaystyle e^{-
    T_u k^+ c_{tot}} (k^+ c_{tot})^M\times \prod_{i=1}^M p(\tau_{b,i}),
\end{equation}
where $T_u=\sum_{i=1}^M \tau_{u,i}$ denotes the total duration for which metallothionein receptors remain unbound during the observation period. The overall concentration of heavy metals and interferer molecules near the receptors is given by $c_{tot}=\sum_{i=1}^N c_i +c_{in}$. The probability $p(\tau_{b,i})$  reflects the likelihood of observing a bound time duration, modeled as a combination of exponential distributions, i.e.,
\begin{equation}
    p(\tau_b)=\sum_{i=1}^N \alpha_i k^-_i e^{-k^-_i \tau_b} + \alpha_{in} k^-_{in} e^{-k^-_{in} \tau_b}.
\end{equation}
The concentration ratios for each heavy metal and interferer molecule are denoted as $\alpha_i=c_i/c_{tot}$ and $\alpha_{in}=c_{in}/c_{tot}$, respectively. The observed unbound and bound time durations are combined into a log-likelihood function defined by \cite{kuscu2022detection}
\begin{equation}
    \begin{split}
            \mathcal{L}(\{\tau_b,\tau_u\}_N) & =ln\,p(\{\tau_b,\tau_u\}_N)   \\
             & =\mathcal{L}_0 + \mathcal{L}(T_u|c_{tot}) + \mathcal{L}(\{\tau_b\}|\boldsymbol{\alpha}),
    \end{split}
\end{equation}
where $\mathcal{L}(T_u|c_{tot})$ and $\mathcal{L}(\{\tau_b\}|\boldsymbol{\alpha})$ denote functions of the total concentration, $c_{tot}$, and the ligand concentration ratios, $\boldsymbol{\alpha}$, respectively. The former indicates that the total unbound time, $T_u$, provides insight into the total ligand concentration, $c_{tot}$, while the latter suggests that the individual bound time durations, $\{\tau_b\}$, reflect the heavy metal concentration ratios, $\boldsymbol{\alpha}$. The term $\mathcal{L}_0$ includes components independent of both $c_{tot}$ and $\boldsymbol{\alpha} = [\alpha_{in}, \alpha_1, \alpha_2, \dots, \alpha_N]^T$. 

We assume that heavy metal concentrations in the reception space behave as a binary CSK modulated signal, allowing us to assess toxicity conditions in the sampling period (i.e., high concentration ratios indicate toxicity (bit-1)). Maximum Likelihood (ML) estimation for ligand concentration ratios lacks an analytical solution, numerical methods are required. \cite{kuscu2022detection} proposes a near-optimal technique by counting binding events within time intervals based on molecule types. For example, the presence of two metal types and one type of interfering molecule in the channel requires three time intervals, with thresholds proportional to the inverse of unbinding rates for lower-affinity molecules, expressed as
\begin{equation}
    T_1=v/k_{in}^- \quad \textrm{and} \quad
    T_2=v/\operatorname{max}(k_1^-,k_2^-),
\end{equation}
where $v > 0$ is the proportionality constant, set to $v = 3$ to minimize estimation error \cite{kuscu2019channel}, determining the likelihood of a ligand binding event within a time range as follows
\begin{equation}
\begin{split}
    p_l&=\int_{T_{l-1}}^{T_{l}} p(\tau_{b})d\tau_b \\ &= \alpha_{in}(e^{-k_{in}^- T_{l-1}}-e^{-k_{in}^- T_{l}}) + \sum_{i=1}^{N} \alpha_i(e^{-k_{i}^- T_{l-1}} - e^{-k_{i}^- T_{l}}). 
\end{split}
\end{equation}
Here, $T_0 = 0$ and $T_{N+1} = +\infty$ \cite{kuscu2022detection}. The probabilities are denoted as $\boldsymbol{p}=\boldsymbol{Q\alpha}$, where $\boldsymbol{p}$ is a probability vector of length $(N+1)$ with elements $p_l$, $\boldsymbol{\alpha}$ is also a vector of length $(N+1)$, and $\boldsymbol{Q}$ is an $(N+1)\times(N+1)$ matrix which is defined as
\begin{equation}
\small
%\begin{split}
   \boldsymbol{Q} = 
    \left(\begin{smallmatrix}
       1-e^{-k_{in}^- T_{1}} & 1-e^{-k_{1}^- T_{1}} &\cdots & 1-e^{-k_{N}^- T_{1}} \\
        e^{-k_{in}^- T_{1}}-e^{-k_{in}^-T_2} & e^{-k_{1}^- T_{1}}-e^{-k_{1}^-T_2} & \cdots & e^{-k_{N}^- T_{1}}-e^{-k_{N}^-T_2} \\
        \vdots & \vdots & \ddots & \vdots  \\
        e^{-k_{in}^- T_{N}} & e^{-k_{1}^- T_{N}}& \cdots & e^{-k_{N}^- T_{N}}  
   \end{smallmatrix}\right).
%\end{split}
\end{equation}

In the context of heavy metal binding to metallothionein proteins, the number of binding events within specific time intervals follows a binomial distribution. 
We assume that $\boldsymbol{n_b}$ consists of elements $n_{b,i}$ that denote the number of binding events within the $i^\text{th}$ time interval, defined by $T_{i-1}$ and $T_i$ and $M$ represents independent binding and unbinding events at equilibrium, with the mean and variance of $\boldsymbol{n_b}$ defined as 
\begin{equation}
    \boldsymbol{\operatorname{E}[n_b}|s_1, \cdots, s_N, n_{in}]=\boldsymbol{p}M,
\end{equation}
\begin{equation}
    \boldsymbol{\operatorname{Var}[n_b}|s_1, \cdots, s_N, n_{in}]=(\boldsymbol{p} \odot (1-\boldsymbol{p}))M,
\end{equation}
where $\boldsymbol{n_b}$ is an $(N+1) \times 1$ vector, $s_i$ denotes the particle number of the $i^\text{th}$ metal type, and $\odot$ is the Hadamard product. 

The Method of Moments (MoM) enables to estimate heavy metal concentration ratios by analyzing binding events within specific time intervals, simplifying the computation compared to ML estimation. Thus, the concentration ratios are derived from the first moment of the distribution, defined as \cite{kuscu2019channel}
\begin{equation}
    \boldsymbol{\hat{\alpha}}=\frac{1}{M}\boldsymbol{Wn_b},
\end{equation}
where $\boldsymbol{W}=\boldsymbol{Q}^{-1}$, i.e., the inverse of $\boldsymbol{Q}$ matrix which is also $(N+1)\times (N+1)$ matrix with elements  $w_{l,i}$. Thus, the estimated concentration ratio of each heavy metal particle and interferer molecule becomes
\begin{equation}
    \hat{\alpha}_l=\frac{1}{M}\sum_{i=1}^{N+1} n_{b,i}w_{l,i}  \quad \text{for }  l \in \{1, \ldots, N, in\}.
    \label{eq:hatalpha_l}
\end{equation}

The variance of the concentration ratio estimator, which reflects the reliability and precision of estimates based on binding events, is expressed as
\begin{equation}
\begin{split}   
\label{eq:variance_estimator}
&\operatorname{Var}[\hat{\alpha}_{l}|s_1, \cdots, s_N, n_{in}]\\
    &=\frac{1}{M^2} \sum_{i=1}^{N+1} \sum_{j=1}^{N+1}w_{l,i}w_{l,j}\operatorname{cov}(n_{b,i}n_{b,j}|s_1, \cdots, s_N, n_{in}),
\end{split}
\end{equation}
where $\l \in \{1, \ldots, N, in\}$ and $\operatorname{cov}(n_{b,i}, n_{b,j})$ is the covariance of binding events across time intervals, defined as
\begin{equation}
    \operatorname{cov}(n_{b,i},n_{b,j})=
   \begin{cases}
   \operatorname{Var}[n_{b,i}|s_1, \cdots, s_N, n_{in}], & \text{if} \ i=j, \\
      -p_ip_jM, & \text{otherwise}.
    \end{cases}
\end{equation}
The covariance matrix, $\boldsymbol{C}$, compactly represents all covariances, defined as $C_{ij} = \operatorname{cov}(n_{b,i}, n_{b,j})$. Here, diagonal elements indicate the variance in binding events within each time interval, while off-diagonal elements reflect relationships between different time intervals. The covariance matrix can also be expressed as $\boldsymbol{C} = M\times\boldsymbol{I_{(N+1) \times (N+1)}} \odot (\boldsymbol{1_{N+1}^\top} \otimes \boldsymbol{p}) - \boldsymbol{p p^\top}$, where $\boldsymbol{p} = \begin{bmatrix} p_1 & p_2 & \dots & p_{N+1} \end{bmatrix}^\top$ denotes the binding event probabilities, and $\otimes$ represents the Kronecker product.
Then, (\ref{eq:variance_estimator}) can be rearranged as
\begin{equation} 
\begin{split}
 &\operatorname{Var}[\hat{\alpha}_l|s_1, \cdots, s_N, n_{in}]  \\  
 &=\frac{1}{M}\left(\left( \sum_{i=1}^{N+1} w_{li}^2 p_i (1 - p_i) \right) - \left( \sum_{i=1}^{N+1} \sum_{\substack{j=1 \\ j \neq i}}^{N+1} w_{li} w_{lj} p_i p_j \right)\right)  \\
 \intertext{Further simplification gives the explicit form,}
 &=\frac{1}{M}\sum_{i=1}^{N+1} \sum_{k=1}^{N+1} w_{li}^2 Q_{ik} \alpha_k \\ 
 &- \frac{1}{M}\sum_{i=1}^{N+1} \sum_{k=1}^{N+1} \sum_{m=1}^{N+1} w_{li}^2 Q_{ik} Q_{im} \alpha_k \alpha_m  \\ 
 &-\frac{1}{M} \sum_{i=1}^{N+1} \sum_{\substack{j=1 \\ j \neq i}}^{N+1} \sum_{k=1}^{N+1} \sum_{m=1}^{N+1} w_{li} Q_{ik} \alpha_k w_{lj} Q_{jm} \alpha_m, 
\end{split}
\end{equation}
where $l \in \{1, \dots, N, in\}$.

The expected value of the concentration ratio estimator reflects the true concentration ratio vector, $\boldsymbol{\alpha}$, providing an unbiased average estimate that closely aligns with actual concentration ratios from binding events, expressed as \cite{kuscu2022detection}
\begin{equation}
\begin{split}
    \boldsymbol{\operatorname{E}[\hat{\alpha}} | s_1, \cdots, s_N, n_{in}] &= \frac{1}{M} \boldsymbol{W} \boldsymbol{\operatorname{E}[n_b} |  s_1, \cdots, s_N, n_{in}] \\
    &= \boldsymbol{Wp} = \boldsymbol{Q}^{-1} \boldsymbol{p} = \boldsymbol{\alpha}.
\end{split}
\end{equation}
The expected value of the individual concentration ratio estimator $\hat{\alpha}_{s_k}$ is given by
\begin{equation}
\operatorname{E}[\hat{\alpha}_{s_k} \mid s_1, \ldots, s_N, n_{in}] = \frac{c_{s_k}}{c_{s_1}  + \cdots + c_{s_N}+c_{in}},
\end{equation}
where $c_{s_k} = s_k/V$ represents the concentration of the $k^\text{th}$ heavy metal molecule, where $s_k$ is its quantity and $V$ is the reception volume. Similarly, $c_{in} = n_{in}/V$, where $n_{in}$ is the number of interferer molecules.

From the perspective of heavy metals and metallothioneins, the law of total expectation and variance simplifies ligand-receptor interactions by breaking down complex calculations. 

Applying the law of total expectation to the concentration ratio estimator, we obtain
\begin{equation}
\label{eq:34}
\begin{split}
\operatorname{E}[\hat{\alpha}_{k}\mid s_k] &=\operatorname{E} \left[ \operatorname{E}[\hat{\alpha}_{s_k}\mid s_1, \ldots, s_N, n_{in}] \mid s_k \right] \\
&= \operatorname{E}\left[\frac{s_k/V}{s_1/V  + \cdots + s_N/V+n_{in}/V} \mid s_k \right].
\end{split}
\end{equation}

Applying the law of total variance to the concentration ratio estimator, we obtain
\begin{equation}
\label{eq:35}
\begin{split}
\operatorname{Var}[\hat{\alpha}_k\mid s_k] &= \operatorname{E}[{\operatorname{Var}}(\hat{\alpha}_k \mid s_1, \ldots, s_N, n_{in}) \mid s_k]  \\ &+ {\operatorname{Var}}[\operatorname{E}(\hat{\alpha}_k \mid s_1, \ldots, s_N, n_{in}) \mid s_k]  \\
&=\frac{1}{M}\sum_{i=1}^{N+1} w_{ki}^2 \sum_{n=1}^{N+1} Q_{in} \operatorname{E}[\hat{\alpha}_n | s_k]  \\ &- \frac{1}{M}\sum_{i=1}^{N+1} w_{ki}^2 \sum_{n=1}^{N+1} \sum_{m=1}^{N+1} Q_{in} Q_{im} \operatorname{E}[\hat{\alpha}_n \hat{\alpha}_m | s_k] \\
&- \frac{1}{M}\sum_{i=1}^{N+1} \sum_{j \neq i} \sum_{n=1}^{N+1} \sum_{l=1}^{N+1} w_{ki} Q_{in} w_{kj} Q_{jl} \operatorname{E}[\hat{\alpha}_n \hat{\alpha}_l | s_k]  \\ &+ \operatorname{Var}[\operatorname{E}(\hat{\alpha}_k \mid s_1, \ldots, s_N, n_{in}) \mid s_k]. 
\end{split}
\end{equation} 
To calculate the conditional expectations in (\ref{eq:34}) and (\ref{eq:35}), we assume that the number of heavy metal and interferer molecules follow a multivariate Gaussian distribution with a known number of a specific heavy metal molecule. This allows us to derive a conditional distribution, where the vector $\boldsymbol{s} = (s_1, \ldots, s_N, n_{in})$ denotes the number of each type of heavy metal and interferer molecule, expressed as
\begin{equation}
    \boldsymbol{s} \sim \mathcal{N}(\boldsymbol{\mu}, \boldsymbol{\Sigma}),
\end{equation}
with mean vector $\boldsymbol{\mu} = (\mu_1, \ldots, \mu_{N}, \mu_{in})$ and covariance matrix $\boldsymbol{\Sigma}$, we assume the number of particular heavy metal, $s_i$, is known, i.e., $s_i = s_i^*$. To derive the conditional distribution of the remaining variables $\boldsymbol{s_{-i}} = (s_1, \ldots, s_{i-1}, s_{i+1}, \ldots, s_{N},n_{in})^\top$ given $s_i = s_i^*$, we partition the mean vector and covariance matrix as follows
\begin{equation}
    \boldsymbol{\mu} = \begin{bmatrix} \mu_i \\ \boldsymbol{\mu_{-i}} \end{bmatrix}, \quad \boldsymbol{\Sigma} = \begin{bmatrix} \Sigma_{ii} & \boldsymbol{\Sigma_{i,-i}} \\ \boldsymbol{\Sigma_{-i,i}} & \boldsymbol{\Sigma_{-i,-i}} \end{bmatrix},
\end{equation}
where $\boldsymbol{\mu_{-i}}$ and $\boldsymbol{\Sigma_{-i,-i}}$ are the mean vector and covariance matrix for the remaining variables, respectively. The mean and variance of $s_i$ are $\mu_i$ and $\Sigma_{i,i}$. $\boldsymbol{\Sigma_{-i,i}}$ and $\boldsymbol{\Sigma_{i,-i}}$ denote the covariances between $s_i$ and $\boldsymbol{s_{-i}}$.

The conditional distribution $\boldsymbol{s_{-i}} \mid (s_i = s_i^*) \sim \mathcal{N}(\boldsymbol{\mu_{-i|s_i}}, \boldsymbol{\Sigma_{s_{-i} | s_i}})$ is also a multivariate normal distribution, with the conditional mean and covariance given by
\begin{equation}
    \boldsymbol{\mu_{s_{-i} | s_i}} = \boldsymbol{\mu_{-i}} + \boldsymbol{\Sigma_{-i,i}} \Sigma_{ii}^{-1} (s_i^* - \mu_i),
\end{equation}
\begin{equation}
    \boldsymbol{\Sigma_{s_{-i} | s_i}} = \boldsymbol{\Sigma_{-i,-i}} - \boldsymbol{\Sigma_{-i,i}} \Sigma_{ii}^{-1} \boldsymbol{\Sigma_{i,-i}}.
\end{equation}

After deriving the conditional distribution from the known joint distribution of heavy metal and interferer molecule numbers, we generate samples using Cholesky decomposition. This method decomposes the positive-definite conditional covariance matrix, $\boldsymbol{\Sigma_{s_{-i} \mid s_i}}$, into the form $\boldsymbol{\Sigma_{s_{-i} \mid s_i}} = \boldsymbol{LL^\top}$. This decomposition simplifies the process of generating correlated Gaussian samples from independent standard normal variables. The generated samples, $\boldsymbol{s_{-i}^*}$, are given by
\begin{equation}
\boldsymbol{s_{-i}^*} = \boldsymbol{\mu_{s_{-i} | s_i}} + \boldsymbol{Lz},
\end{equation}
where $\boldsymbol{z} \sim \mathcal{N}(\boldsymbol{0, I})$ is a vector of standard normal variables. These samples enable calculation of the mean and variance of the concentration ratio distribution in (\ref{eq:34}) and (\ref{eq:35}). Although the concentration ratio may not be Gaussian, simulations show close alignment. The histogram and probability density function based on the sample mean and standard deviation strongly agree, justifying the use of the complementary error function (erfc) for Gaussian-like data.

Assuming that heavy metal concentrations in the bloodstream behave like a binary CSK modulated signal with toxicity information encoded in their concentrations, we define the variable $x_k$ to represent the toxicity information. This is detected by estimating the concentration ratio of the $k^{\text{th}}$ heavy metal during the sampling period, leading to a decision of 0 (non-toxic) or 1 (toxic). The decision rule is then defined as
\begin{equation}
    \hat{x} = \underset{x_k \in \{0,1\}}{\arg \max} \, p(\kappa | x_k),
\end{equation}
where $\kappa$ represents the estimated concentration ratio of a specific heavy metal, serving as the received signal statistic used to detect toxicity conditions. The decision rule is simplified with a threshold $\lambda$, expressed as $\kappa \underset{H_0}{\overset{H_1}{\gtrless}} \lambda$. The optimal threshold $\lambda$, minimizing error probability for normally distributed statistics, is given by \cite{kuscu2022detection}
\begin{equation}
\small
\begin{split}
\lambda &= \gamma^{-1}\\& \times \left( \frac{\operatorname{Var}[\kappa | x_k=1] \operatorname{E}[\kappa | x_k=0] - \operatorname{Var}[\kappa | x_k=0] \operatorname{E}[\kappa | x_k=1]}{\operatorname{Var}[\kappa | x_k=1] - \operatorname{Var}[\kappa | x_k=0]} \right) \\
&+ \operatorname{Std}[\kappa | x_k=1] \operatorname{Std}[\kappa | x_k=0] \\ &\times \left( \frac{(\operatorname{E}[\kappa|x_k=1] - \operatorname{E}[\kappa|x_k=0])^2}{2 \gamma \ln \left( \frac{\operatorname{Std}[\kappa|x_k=1]}{\operatorname{Std}[\kappa|x_k=0]} \right)} + 2 \gamma \ln \left( \frac{\operatorname{Std}[\kappa|x_k=1]}{\operatorname{Std}[\kappa|x_k=0]} \right) \right),
\end{split}
\end{equation}
where $\gamma = \operatorname{Var}[\kappa|x_k=1] - \operatorname{Var}[\kappa|x_k=0]$ denotes the variance difference in detecting toxicity condition of heavy metal concentration ratios. The standard deviation, $Std[\cdot] = \sqrt{\operatorname{Var}[\cdot]}$, is then calculated. With the threshold set, the detection performance is evaluated using the BEP, expressed as
\begin{flalign*}
    p_k(\epsilon) = \frac{1}{2} \left[ p_k (\hat{x}_k = 1 | x_k= 0) + p_k (\hat{x}_k = 0 | x_k = 1) \right]
\end{flalign*}
\begin{equation}
\small
    = \frac{1}{4} \left[ erfc \left( \frac{\lambda - \operatorname{E}[\kappa|x_k=0]}{\sqrt{2 \operatorname{Var}[\kappa|x_k=0]}} \right) + erfc \left( \frac{\operatorname{E}[\kappa|x_k=1] - \lambda}{\sqrt{2 \operatorname{Var}[\kappa|x_k=1]}} \right) \right].
\end{equation}

\subsection{Mathematical Model of PaCoT Release Mechanism}
Each toxic particle in the PaCoT diffuses freely based on Fick's laws. In one dimension, the probability density function of a Brownian particle satisfies the diffusion equation \cite{bian2016111}.
\begin{equation}
\frac{\partial f(x, t)}{\partial t} = D \frac{\partial^2 f(x, t)}{\partial x^2},
\end{equation}
where $D$ is the diffusion coefficient, given by $D = \frac{k_B T}{6 \pi \mu_{eff} r}$, with $k_B$ as Boltzmann's constant, $T$ as temperature, $r$ as the particle radius, and $\mu_{eff}$ as the effective viscosity, defined as $\mu_{eff} = \mu_w + \sum_{i=1}^{N} \phi_i \mu_i$, where $\mu_w$ is the water viscosity and $\phi_i \mu_i$ represents the contribution from each molecule’s volume fraction, $\phi_i$ and viscosity. The individual viscosity, $\mu_i$, is defined as $\mu_i = \frac{(m_i k_B T)^{1/2}}{\pi^{3/2} d_i^2}$, where $m_i$ is molecular mass, and $d_i$ is particle diameter of the $i^{\text{th}}$ heavy metal molecule. Higher viscosity lowers $D$, which slows particle release. The probability density function $f(x, t)$ describes the particle position over time, following a Gaussian distribution with zero mean and variance $2Dt$, expressed as $f(x, t) = \frac{1}{\sqrt{4 \pi D t}} e^{-\frac{x^2}{4Dt}}$. 

This study designs a single-disc viscous micropump in the PaCoT to enable rapid particle release, influenced by C-shaped channel geometry (e.g., height, width, radius) and boundary conditions. These factors create a pressure difference, $\Delta P$, that directs toxic molecules to the release point. The volumetric flow rate, $Q$, governs particle transport speed, with higher $Q$ indicating faster release. It is given by \cite{blanchard2006miniature}
\begin{equation}
    Q  = \frac{h^3 \ln{(R_{1}/R_{2})}}{12\mu_{eff}} \frac{\Delta P}{\Delta \theta} + \frac{\omega h(R_{2}^2-R_{1}^2)}{4},
    \label{eq:quadratic_flow_rate}
\end{equation}
where $h$ is C-shaped channel height, while $R_1$ and $R_2$ denote the inner and outer radii of the C-shaped channel. $\Delta \theta$ is the circumferential angle over which the pressure acts, $\omega$ is the rotational speed (in revolutions per minute (RPM)). This expression indicates that the flow rate is directly proportional to the pressure difference and rotational speed, both crucial for controlling the particle release rate.
For simplicity, (\ref{eq:quadratic_flow_rate}) assumes a negligible aspect ratio for the C-shaped channel and disregards gravity, as low disc speeds minimize centrifugal forces, allowing viscous and pressure forces to dominate.

Experimental studies \cite{al2007investigation, alparametric} have shown that changing the aspect ratio leads to deviations from expected results \cite{al2009recent}. Reduced channel height, higher rotational speed, and increased fluid viscosity enhance pressure rise but reduce flow rates in highly viscous mixtures. These studies separately examined radius ratio and aspect ratio effects on micropump flow performance. A generalized volumetric flow rate model is developed to estimate the combined influence of geometric parameters and boundary conditions on performance, expressed as
\begin{equation}
    Q_{general} = \frac{1}{2} F_{DR}\, F_{DA} - \frac{1}{12} \overline{\operatorname{Re}} \, \operatorname{Eu} \, F_{PR}\, F_{PA}, 
    \label{Q_general}
\end{equation}
where $\overline{\operatorname{Re}} = \frac{\rho \overline{U} h^2}{R_m \Delta \theta^2 \mu}$ represents the reduced Reynolds number, $\operatorname{Eu} = \frac{\Delta p}{\rho \overline{U}^2}$ denotes the Euler number, and $R_m = (R_1 + R_2)/2$ is the average radius. Here, $F_{DR}$ and $F_{PR}$ are drag and pressure shape factors for the radius ratio, while $F_{DA}$ and $F_{PA}$ correspond to the aspect ratio, defined as \cite{al2009recent}
\begin{align}
    F_{DA} &= \frac{4 \, \left(\frac{w}{h}\right)}{\pi^3} \sum_{n=1}^{\infty} \frac{((-1)^n - 1)^2 \cosh(\frac{n\pi h}{w}) - 1}{n^3 \sinh(\frac{n\pi h}{w})}, \\
    F_{DR} &= 1.0, \\
    F_{PA} &= \frac{\left(\frac{w}{h}\right)^2}{\pi^5} - \frac{48 \, \left(\frac{w}{h}\right)^3}{\pi^5} \sum_{n=1}^{\infty} \frac{((-1)^n - 1)^2 \cosh(\frac{n\pi h}{w}) - 1}{n^5 \sinh(\frac{n\pi h}{w})}, \\
    F_{PR} &= \frac{1}{2} \left( \frac{\left(\frac{R_1}{R_2}\right) + 1}{\left(\frac{R_1}{R_2}\right) - 1} \right) \ln\left( \frac{R_1}{R_2} \right),
\end{align}
where $w$ represents the C-shaped channel width. When calculating shape factors, the radius ratio approaches one, ($R_1 / R_2 \rightarrow 1$), when determining the drag and pressure shape factors based on aspect ratio values. Similarly, the aspect ratio approaches zero, ($h / w \rightarrow 0$), when determining the drag and pressure shape factors based on radius ratio values \cite{al2007investigation}.

The release mechanism, combining the micropump flow rate and molecular diffusion, is described by a one-dimensional advection-diffusion equation. This study assumes a Gaussian distribution for the concentration of collected toxic heavy metal particles in the PaCoT. The rearranged advection-diffusion equation is expressed as
\begin{equation}
    c(x,t)=\frac{M}{\sqrt{4\pi t}}e^{-\frac{(x-ut)^2}{4Dt}},
\end{equation}
where $c(x,t)$ denotes the concentration of collected toxic metal particles in the PaCoT at position $x$ and time $t$, where $D$ is the diffusion coefficient, $u$ is the average flow velocity along the x-axis generated by the single-disc viscous micropump, and $M$ denotes the initial concentration of the toxic particle. Over time, the concentration evolves due to advection and diffusion, with a Gaussian term describing its spreading.

\subsection{Energy Consumption}
This study evaluates system energy consumption by focusing on the energy required for the release mechanisms driven by the micropump in the PaCoT, while disregarding the energy consumption associated with the reception process.

Energy consumption is a critical factor in the operation of the single-disc viscous micropump, which controls the release of collected heavy metal particles from the PaCoT. The energy required by the micropump depends on parameters such as rotational speed, pressure gradient, and the C-shaped channel geometry, affecting the volumetric flow rate and pressure difference. Thus, it can be expressed as
\begin{equation}
    E = \Delta P \times Q \times \Delta t,
    \label{eq:micropump_energy_consumption}
\end{equation}
where $\Delta t$ is the time difference, $\Delta P$ is the pressure difference and $Q$ is the volumetric flow rate in the C-shaped micropump channel. $\Delta P$ is defined as
\begin{equation}
    \Delta P = \frac{6 \mu_{eff} V \Delta L}{h^2},
\end{equation}
where $\mu_{eff}$ is the effective viscosity of the suspension, $V = \omega (R_1 + R_2)/2$ is the average velocity, and $\Delta L = \Delta \theta (R_2 + R_1)/2$ is the flow path length of the micropump.

\section{Performance Analysis} \label{section4}
This section presents numerical results evaluating the PaCoT's reception and release mechanisms under various configurations. The reception side of PaCoT functions as a receiver, estimating the concentration ratio of specific heavy metals, which is treated as a received signal. The threshold determines the toxicity condition, with a high concentration ratio indicating a toxic state (detected as bit-1) and a low ratio indicating a safe state (detected as bit-0). We assume that binary CSK modulation encodes toxicity information into heavy metal concentrations in the MC channel (e.g., blood vessel) where high and low levels serve as transmitted bits representing true toxic (bit-1) and safe (bit-0) conditions, respectively. Detection accuracy is assessed using BEP, where an error occurs whenever the detected toxicity differs from the actual condition. A lower BEP indicates higher accuracy, so selecting an appropriate threshold is essential to minimize detection errors. The release time is another critical metric that measures the efficiency of the transport of toxic particles to the release node. It is influenced by both passive and active mechanisms. Energy consumption reflects energy efficiency, indicating the system's potential for long-term operation and continuous monitoring. These metrics assess the PaCoT’s performance in detecting, collecting, and transmitting toxic heavy metals. Parameters used are listed in Tables \ref{table1} and \ref{table2}.

Our performance analysis examines two scenarios involving molecule reception and release mechanisms. For the reception mechanism, we consider three molecule types: Zn, Cd, and an interferer. Binding rates and dissociation constants for Zn and Cd are based on physiological conditions at pH $7.4$ with $5$ \textmu$\unit{M}$ apo-MT and $35$ \textmu$\unit{M}$ of Zn and Cd \cite{korkola2024structural}, enhancing the model's realism. The unbinding rate for the interferer is estimated by dividing the maximum unbinding rate of Zn and Cd by the affinity ratio. We assume a low correlation ($0.2$) between the interferer and metal ions, with a coefficient of variation ($0.1$) for all molecule types. In the release mechanism, molecules collected at the reception side move through PaCoT’s internal structure, modeled as a stochastic process with directional bias. This movement involves advection driven by the micropump’s fluid flow and diffusion due to concentration gradients. As molecular concentration rises, collisions and dispersion increase, modeled as Gaussian noise to capture randomness.
\begin{table}[h]
\scriptsize % Reduces the font size to script size
    \centering
    \captionsetup{
        format=myformat1, 
        labelsep=none, % No separator between label and caption
        justification=centering, % Ensure caption text is centered
    }
    \caption{\textsc{Simulation Parameters}}
    \label{table2}
    
    \begin{tabular}{l|l} % Align columns to the left with 'l'
    
    \textbf{Parameter, Symbol, Reference} & \textbf{Value}\\
    \hline
    \textit{Boltzman constant,} ($k_B$) & $1.3807\times10^{-14}$ \unit{\frac{mm^2g}{Ks^2}} \\
    \hline
    \textit{Temperature,} ($T$) & $300$ \unit{K}  \\
    \hline
    \textit{Water viscosity,} ($\mu_w$) & $0.0008509$ \unit{\frac{g}{mm(s)}} \\
    \hline
    \textit{Water density,} ($\rho_w$) & $996.64\times10^{-6}$ \unit{\frac{g}{mm^3}}  \\
    \hline
    \textit{Volume fraction of water,} ($\phi_w$) & 0.9 \\
    \hline
    \textit{Radius of PaCoT,} ($R$) & $3$ \unit{mm}  \\
   \hline
   \textit{Length of PaCoT,} ($h$) & $12$ \unit{mm}  \\
   \hline
   \textit{Inner radius of C-shaped channel,} ($R_1$), \cite{blanchard2006miniature}  &  $1.19$ \unit{mm}  \\
   \hline
   \textit{Outer radius of C-shaped channel,} ($R_2$), \cite{blanchard2006miniature} & $2.38$ \unit{mm} \\
   \hline
   \textit{Angular velocity of micropump,} ($w$), \cite{blanchard2005single} & $500$ \unit{rpm} \\
   \hline
   \textit{Binding rate of Cd molecule,} ($k^+_{cd}$), \cite{korkola2024structural} & $5.7 \times 10^7$ \unit{M^{-1}s^{-1}} \\
   \hline 
   \textit{Dissociation constant of Cd,} ($Kd_{cd}$), \cite{korkola2024structural} & $5.1 \times 10^{-6}$ \unit{M}\\
   \hline 
   \textit{Binding rate of Zn  molecule,} ($k^+_{zn}$), \cite{korkola2024structural}& $5.1 \times 10^7$ \unit{M^{-1}s^{-1}} \\
   \hline 
   \textit{Dissociation constant of Zn,} ($Kd_{zn}$), \cite{korkola2024structural} & $6 \times 10^{-6}$ \unit{M}\\
   \hline 
   \textit{Affinity ratio,} ($\eta$), \cite{kuscu2022detection} & $0.2$  \\
   \hline 
   \textit{Binding rate of interferer molecule,} ($k^+_{in}$), \cite{korkola2024structural} & $4 \times 10^{7}$ \unit{M^{-1}s^{-1}}\\
   \hline 
   \textit{Reception space,} ($V$), \cite{kuscu2022detection} & $4\times 10^{-12}$ \unit{L} \\
   \hline 
   \textit{Mean number of interferer molecule,} ($\mu_{in}$), \cite{kuscu2022detection} &  $2 \times Kd_{in}V$  \\ 
    \hline 
   \textit{Mean number of Cd molecule,} ($\mu_{cd}$), \cite{kuscu2022detection}& $3 \times Kd_{cd}V$  \\ 
   \hline 
   \textit{Concentration of Zn molecule,} ($c_{x_{zn}=0}$),   \cite{kuscu2022detection} & $4 \times Kd_{zn}$  \\ 
   \hline 
   \textit{Concentration of Zn molecule,} ($c_{x_{zn}=1}$),   \cite{kuscu2022detection} & $5 \times Kd_{zn}$  \\ 
   \hline 
   \textit{Correlation between interferer and Cd,} ($\rho_{in-cd}$) & $0.2$  \\ 
   \hline 
   \textit{Correlation between interferer and Zn,} ($\rho_{in-zn}$) & $0.2$  \\ 
   \hline 
   \textit{Correlation between Zn and Cd,} ($\rho_{zn-cd}$), \cite{balfourier2022importance} & $0.7$  \\ 
   \hline 
    \end{tabular}
\end{table}

\begin{figure*}[h]
\captionsetup{justification=centering, margin=0.1cm}  
\begin{subfigure}{0.33\textwidth}
\includegraphics[scale=0.17]{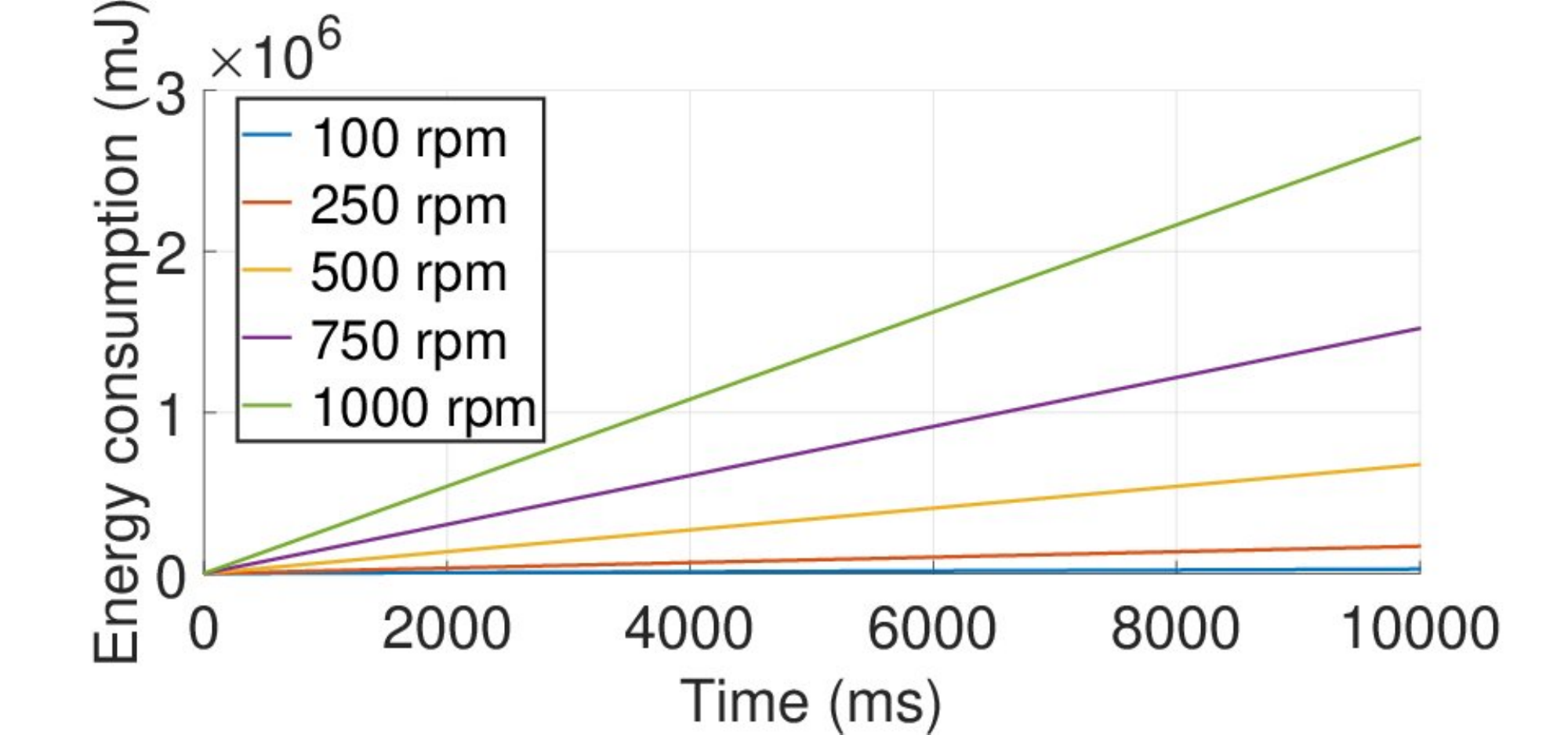} 
\caption{}
\label{fig:subim1}
\end{subfigure}
\begin{subfigure}{0.33\textwidth}
\includegraphics[scale=0.17]{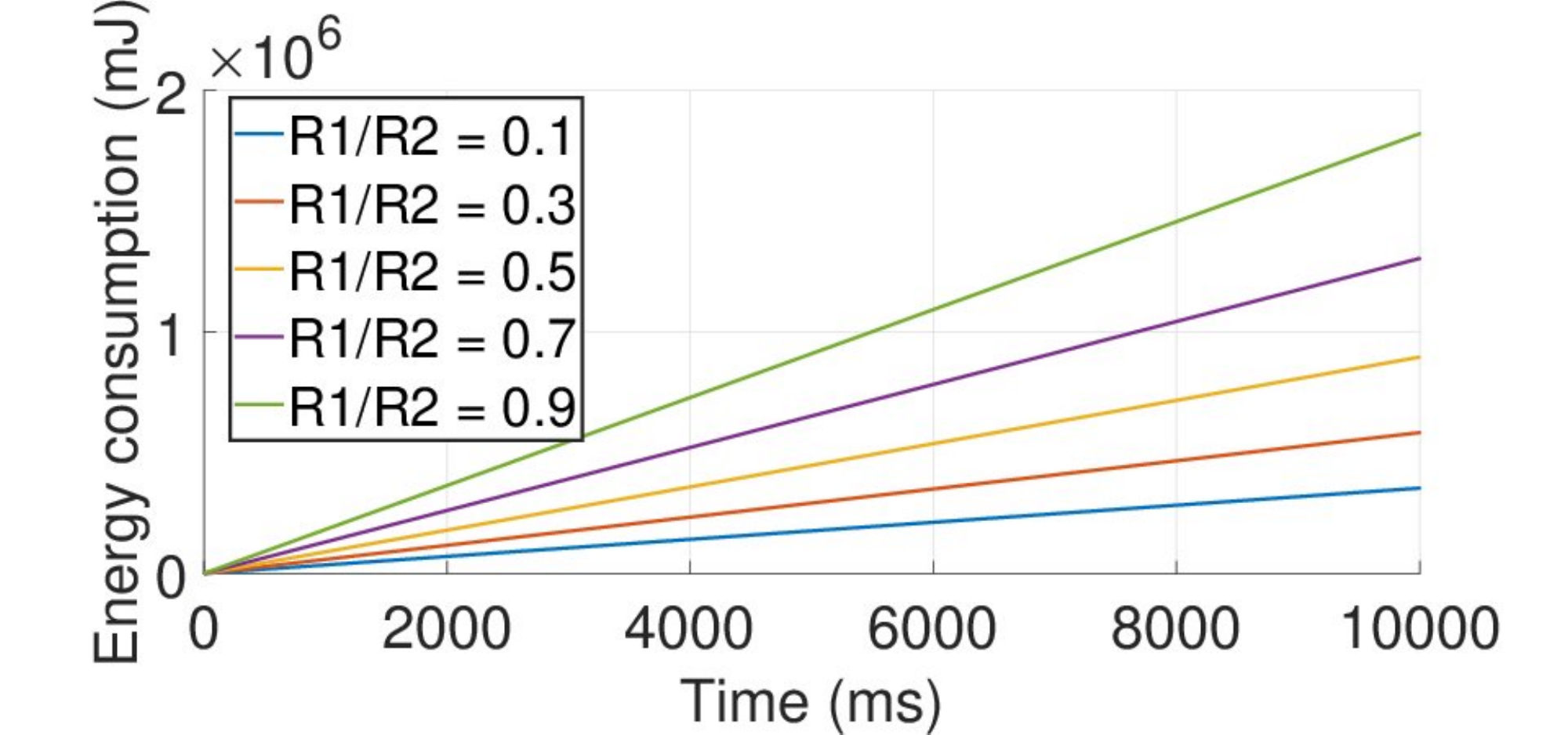} 
\caption{}
\label{fig:subim12}
\end{subfigure}
\begin{subfigure}{0.33\textwidth}
\includegraphics[scale=0.17]{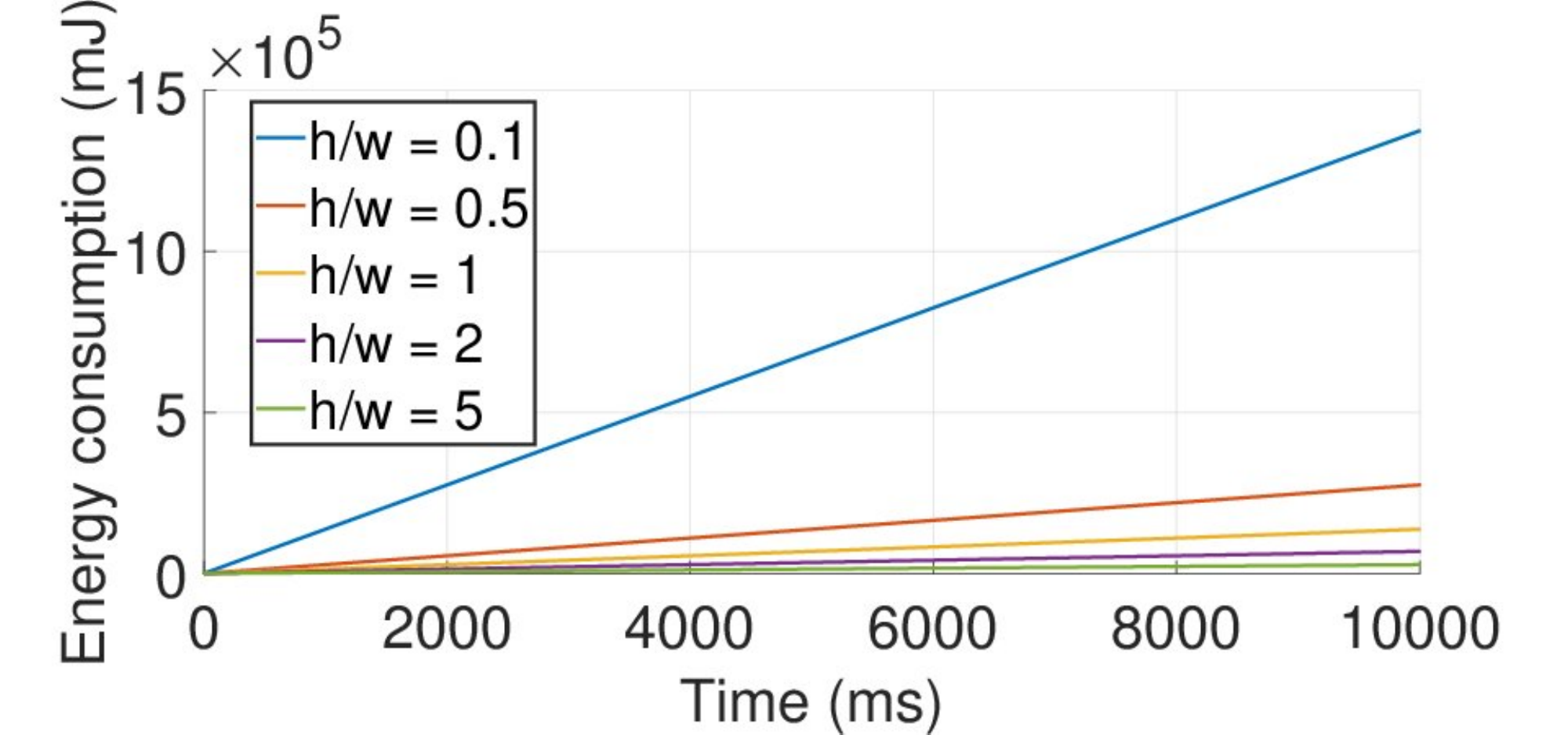}
\caption{}
\label{fig:subim2}
\end{subfigure}
\caption{.\hspace{0.1 cm} Energy consumption of the micropump over time vs. (a) RPM, (b) radius ratio, and (c) aspect ratio.}
\label{fig:microPumpEnergyConsumption}
\end{figure*}

The theoretical framework for micropump energy consumption is based on the power required to maintain fluid flow, determined by the pressure differential and volumetric flow rate. Fig. \ref{fig:microPumpEnergyConsumption} shows energy consumption trends over time with varying RPM values, radius ratios, and aspect ratios. The results indicate a linear increase in energy consumption with higher RPM values, larger radius ratios, and lower aspect ratios, aligning with theoretical predictions. Calculations use the effective viscosity values of molecules in Table \ref{table1}, with the inner and outer radii of the single-disc viscous micropump set to $1.19$ \unit{mm} and $2.38$ \unit{mm} \cite{blanchard2006miniature}. The angular span in the C-shaped channel, denoted by $\Delta \theta$, is set to $\pi/2$ \cite{blanchard2006miniature}. Fig. \ref{fig:microPumpEnergyConsumption}(a) contrasts with Fig. \ref{fig:microPumpEnergyConsumption}(b), where energy consumption changes with inner radius, keeping RPM at $500$ and outer radius fixed. Fig. \ref{fig:microPumpEnergyConsumption}(c) sets the radius ratio at $0.9$ and RPM at $500$, validating assumptions of steady flow and consistent pressure, confirming no significant non-linearities or turbulence in release process.

\begin{figure}[h]
\captionsetup{labelformat=myformat}  
\begin{subfigure}{0.5\textwidth}
\includegraphics[scale=0.28]{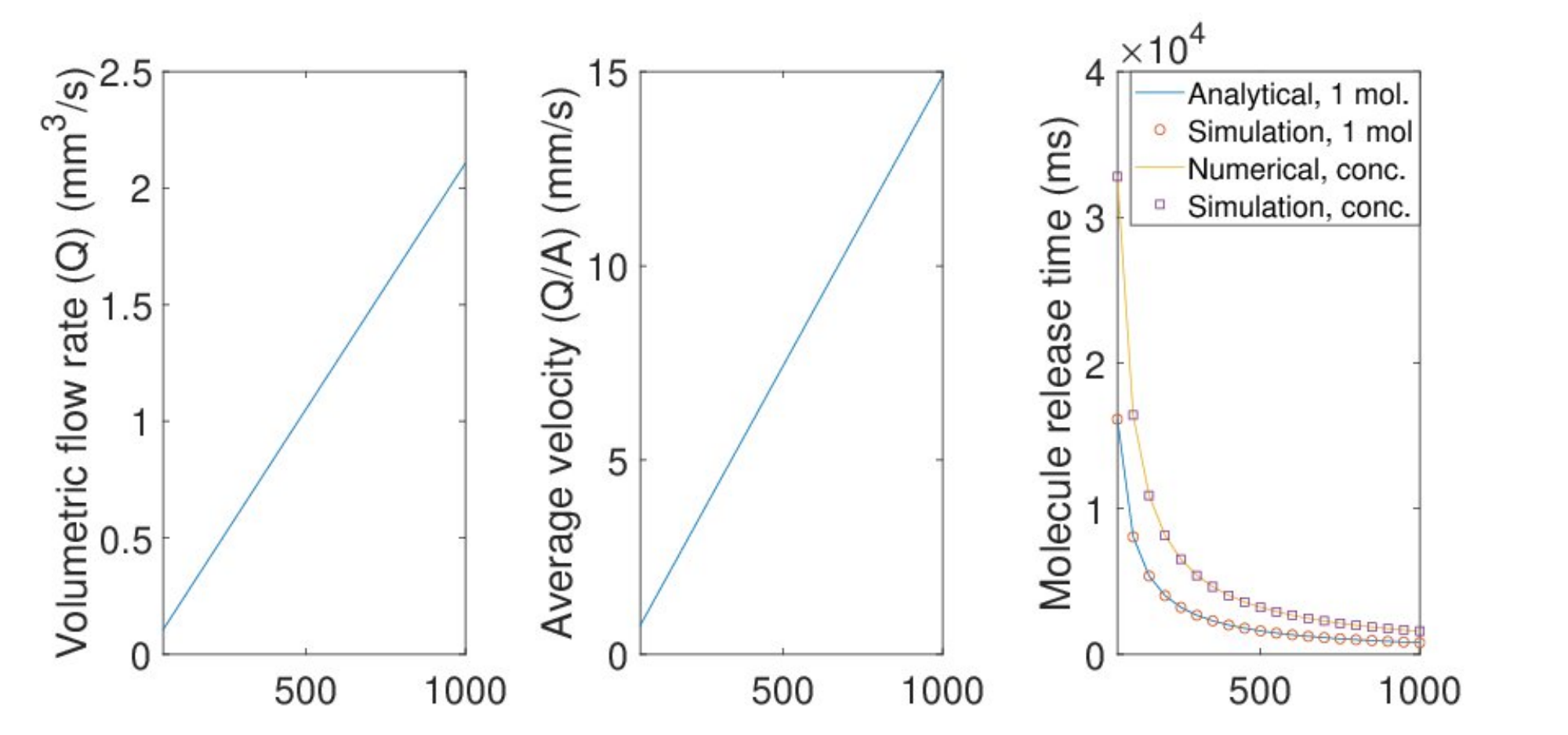} 
\caption{\centering}
\label{fig:subim1}
\end{subfigure}
\quad
\begin{subfigure}{0.5\textwidth}
\includegraphics[scale=0.28]{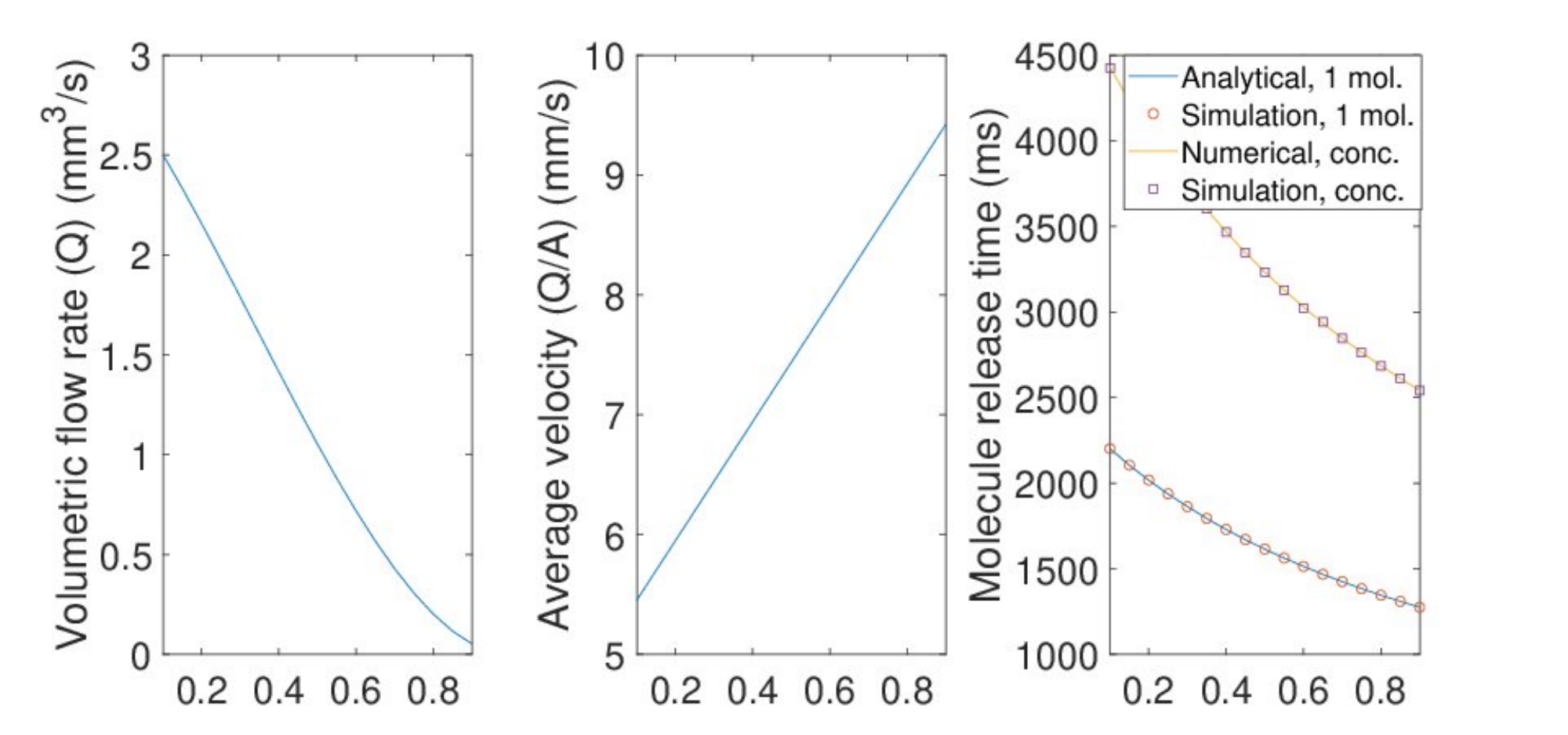} 
\caption{\centering}
\label{fig:subim12}
\end{subfigure}
\quad
\begin{subfigure}{0.5\textwidth}
\includegraphics[scale=0.268]{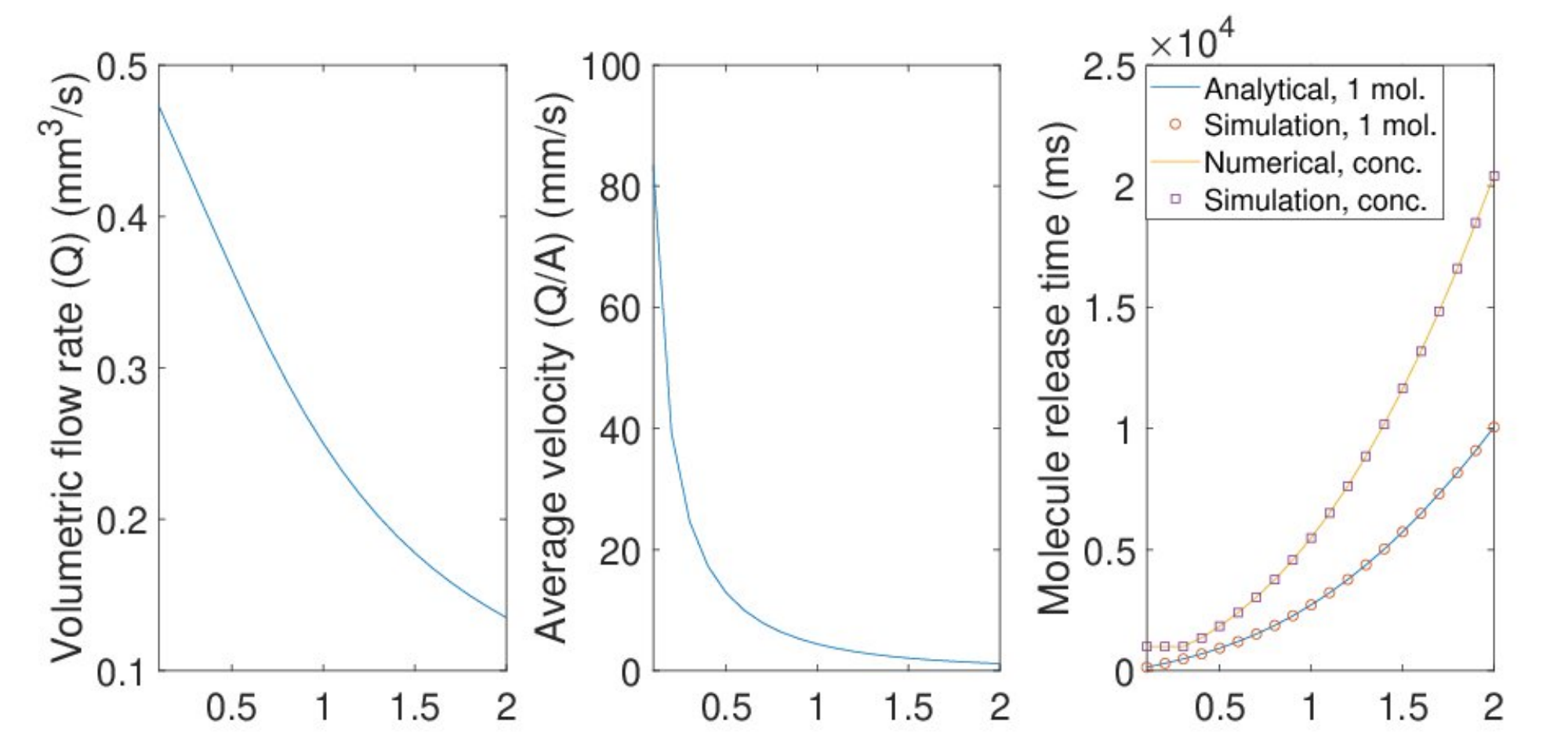}
\caption{\centering}
\label{fig:subim2}
\end{subfigure}
\quad
\caption{\hspace{0.1 cm} Volumetric flow rate, average velocity, and molecule release time from PaCoT vs. (a) RPM, (b) radius ratio ($R_1/R_2$), and (c) aspect ratio.}

\label{fig:ReleaseTime}  
\end{figure}

Fig. \ref{fig:ReleaseTime} illustrates how the micropump's RPM, aspect ratio, and radius ratio affect flow rate, velocity, and molecule release time. These parameters are critical for understanding PaCoT’s molecule transport dynamics, where diffusion and advection are key. Volumetric flow rate is calculated using (\ref{eq:quadratic_flow_rate}) in Fig. \ref{fig:ReleaseTime}(a) and \ref{fig:ReleaseTime}(b), and (\ref{Q_general}) in Fig. \ref{fig:ReleaseTime}(c), assuming maximum flow rate and minimal pressure difference ($\Delta P = 0.0$).

\subsection{Impact of RPM on Molecule Release Time} Fig. \ref{fig:ReleaseTime}(a) shows results with inner radius $R_1 = 1.19$ \unit{mm} and outer radius $R_2 = 2.38$ \unit{mm}, yielding an aspect ratio of $0.1$. RPM values range from $50$ to $1000$ in steps of $50$. The first two graphs in Fig. \ref{fig:ReleaseTime}(a) show that volumetric flow rate and average flow velocity increase linearly with rising RPM, indicating enhanced advection and improved molecule transport. Molecule release time starts above $35$ \unit{s} at low RPM and drops to about $3$ \unit{s} at $1000$ RPM due to enhanced convective flow. Simulation and analytical curves for single molecules align closely, indicating advection dominance, while total concentration shows slightly higher values due to diffusion effects. Increased RPM accelerates advection, reducing release time and allowing molecules to move more quickly, as shown in the third graph of Fig. \ref{fig:ReleaseTime}(a).
\subsection{Impact of Radius Ratio on Molecule Release Time} Fig. \ref{fig:ReleaseTime}(b) analyzes the effect of varying the radius ratio at an RPM of $500$ and an aspect ratio of $0.1$, with $R_2$ held constant and $R_1$ varying, resulting in a radius ratio range from $0.1$ to $0.96$, increasing in steps of $0.05$. The first plot in Fig. \ref{fig:ReleaseTime}(b) shows a decrease in volumetric flow rate as the radius ratio increases, which is expected due to the shrinking flow area as $R_1$ nears $R_2$, while the second plot reveals a linear increase in average flow velocity, reflecting mass conservation in the system. The third plot in Fig. \ref{fig:ReleaseTime}(b) compares analytical and simulation results for molecule release time, showing a decrease in release time with increasing radius ratio and strong agreement between the analytical and simulation methods. These findings confirm that higher radius ratios, while reducing flow rate, increase flow velocity and shorten molecule transport time, aligning with theoretical calculations.  
\subsection{Impact of Aspect Ratio on Molecule Release Time} Fig. \ref{fig:ReleaseTime}(c) examines the effect of aspect ratio on micropump performance, varying from $0.1$ to $2$ in increments of $0.1$, with a fixed radius ratio $(R_1/R_2=0.9)$ and an RPM of $500$. The first plot in Fig. \ref{fig:ReleaseTime}(c) shows a decline in volumetric flow rate as the aspect ratio increases, due to the narrowing flow channel that restricts fluid movement. The second plot in Fig. \ref{fig:ReleaseTime}(c) indicates a significant decrease in average flow velocity across the cross-section with rising aspect ratio, highlighting the inverse relationship between aspect ratio and flow velocity. The third plot in Fig. \ref{fig:ReleaseTime}(c) reveals that molecule release time grows exponentially with increasing aspect ratio, with strong agreement between analytical and simulation results. Thus, larger aspect ratios result in slower molecular transport, as expected due to increased flow resistance in narrower channels. 

\begin{figure}[htp]
    \centering        \includegraphics[scale=0.34]{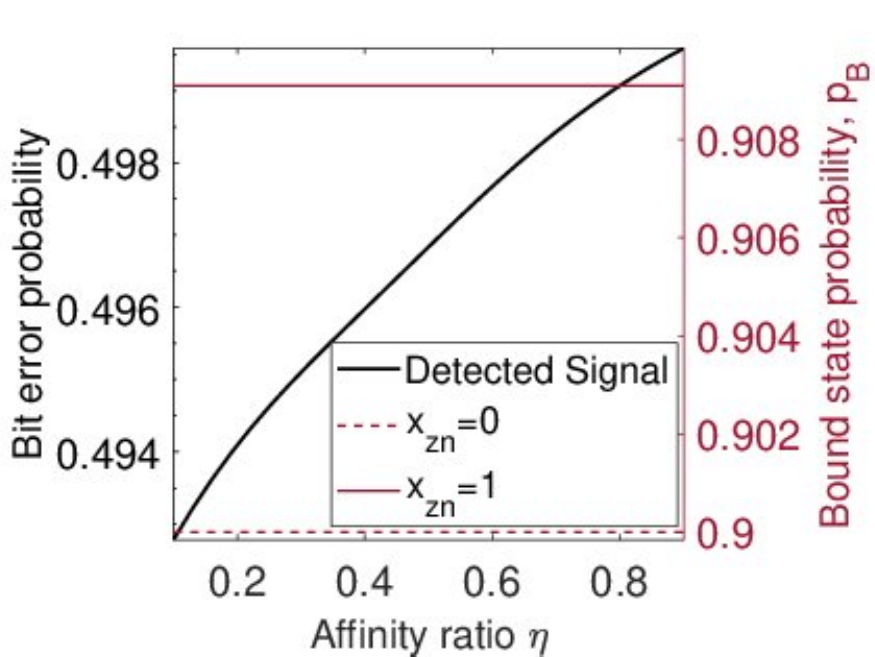}    \captionsetup{labelformat=myformat}
    \caption{\hspace{0.1 cm} BEP and bound state probability vs. affinity ratio variation.}
    \label{fig:BER_affinity}   
\end{figure}
The affinity ratio determines the similarity between heavy metal molecules and interferer molecules. To assess its impact on detection performance, we consider $\eta<1$. In this simulation, the mean concentrations of Cd and the interferer molecule, as well as the concentration of Zn and the unbinding rates for Zn and Cd, are held constant, as shown in Table \ref{table2}. However, the unbinding rate of the interferer molecule, $k_{in}^-$, varies according to the affinity ratio. Fig. \ref{fig:BER_affinity} illustrates the relationship between the affinity ratio and both the BEP and the bound state probability. The BEP curve shows a slight increase with a higher affinity ratio, indicating a marginal rise in toxicity detection errors. In contrast, the bound state probabilities for both $x_{Zn}=0$ and $x_{Zn}=1$ remain relatively stable, suggesting that changes in the affinity ratio have minimal impact on bound state probabilities.
\begin{figure}[htbp]
\captionsetup{labelformat=myformat}  
\centering
\begin{subfigure}{0.49\linewidth} % Each subfigure takes less than half the column width
\centering
\includegraphics[scale=0.23]{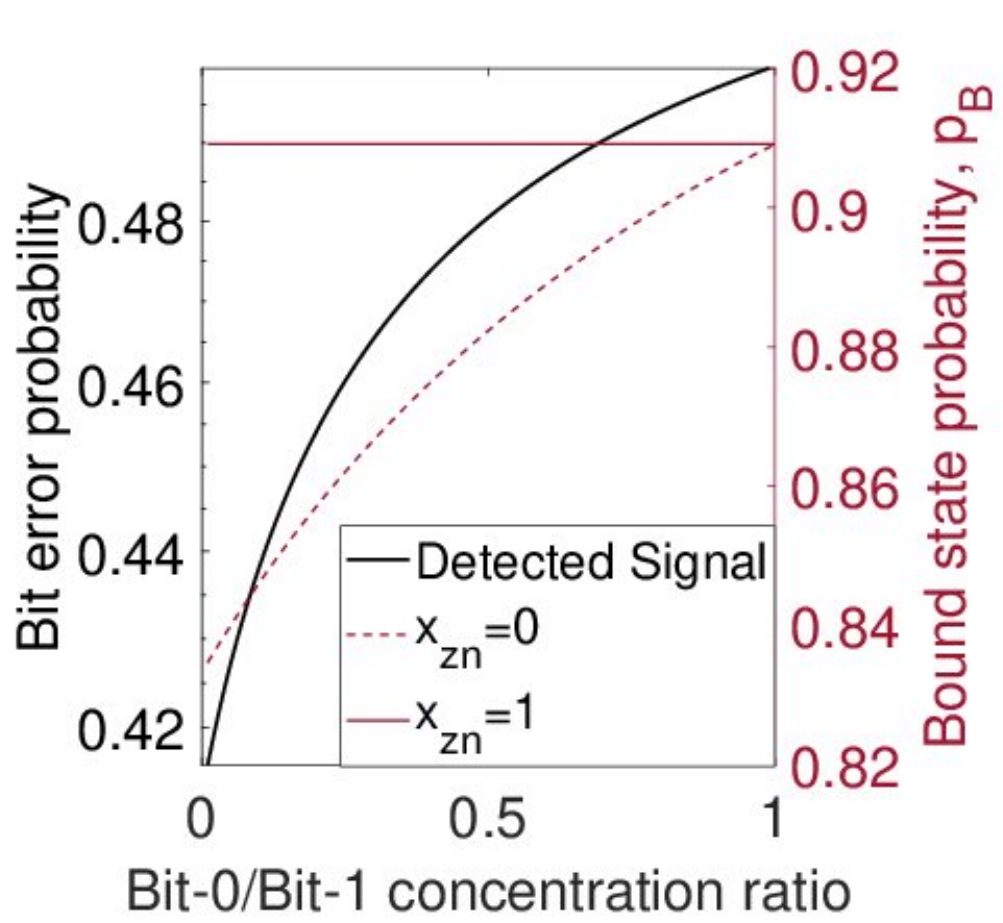} 
\caption{\centering}
\end{subfigure}
\hfill % Adds horizontal spacing between subfigures
\begin{subfigure}
{0.49\linewidth} % Same width as the first subfigure
\centering
\includegraphics[scale=0.23]{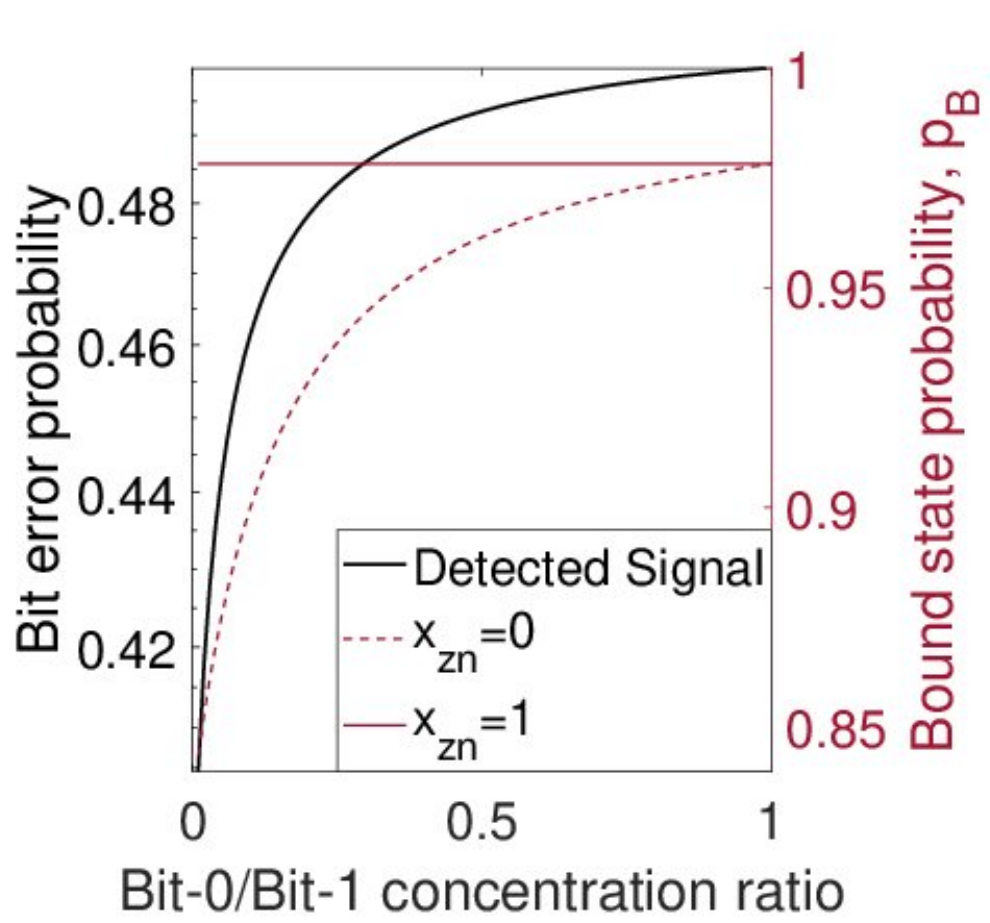} 
\caption{\centering}
\end{subfigure}
\caption{\hspace{0.1 cm} BEP and bound state probability vs. bit-0/bit-1 concentration ratio under (a) non-saturation and (b) saturation conditions.}
\label{fig:BER_bit} 
\end{figure}

Fig. \ref{fig:BER_bit} shows the relationship between BEP and bound state probability in ligand-receptor binding under non-saturation $(c_{x_{zn}=0}=4\times Kd_{zn} \text{ and } c_{x_{zn}=1}=5\times Kd_{zn})$ and saturation $(c_{x_{zn}=0}=39\times Kd_{zn} \text{ and } c_{x_{zn}=1}=40\times Kd_{zn})$ conditions. We examine the effect of varying the ratio of heavy metal concentrations for safe condition (bit-0) and toxic condition (bit-1), ranging from $0.1$ to $0.99$. The results indicate that in non-saturation, the detection method performs significantly better when concentration levels are well-separated (low ratio). As the concentration ratio increases, detection performance declines, especially under saturation, where BEP and the bound state probability for bit-0 rise more steeply. Overall, saturation leads to a higher BEP, while non-saturation results in a more gradual increase, indicating better detection performance in non-saturated conditions.

\begin{figure*}[htbp]
\captionsetup{justification=centering, margin=0.1cm}  
\begin{subfigure}{0.33\textwidth}
\centering
\includegraphics[scale=0.34]{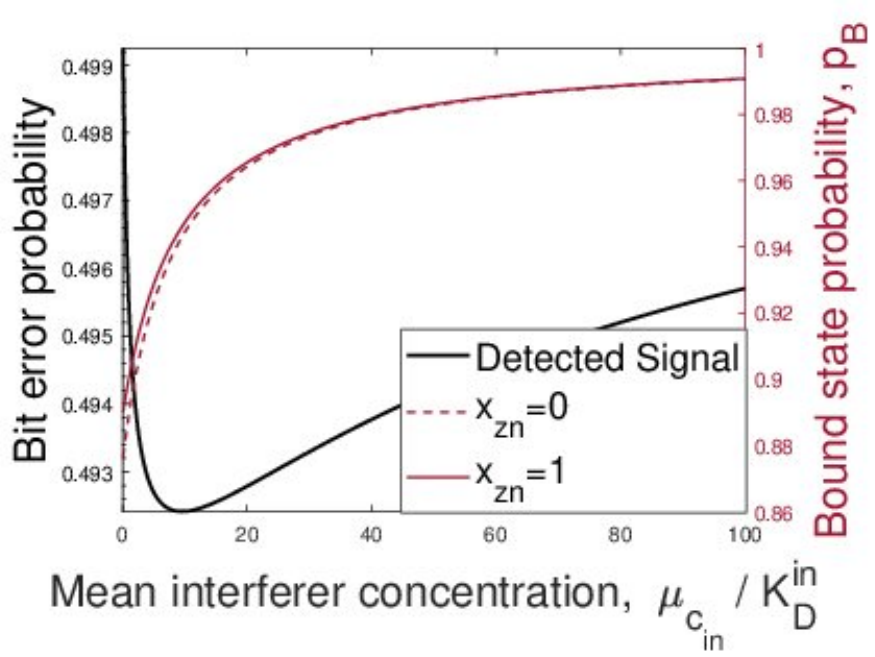} 
\caption{}
\end{subfigure}
\begin{subfigure}{0.33\textwidth}
\centering
\includegraphics[scale=0.34]{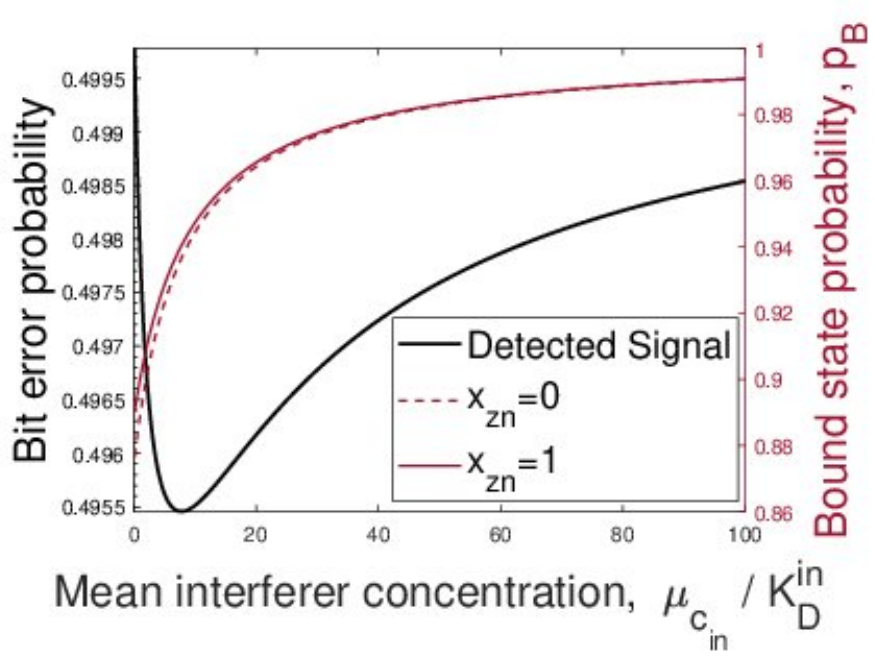} 
\caption{}
\end{subfigure}
\begin{subfigure}{0.33\textwidth}
\centering
\includegraphics[scale=0.34]{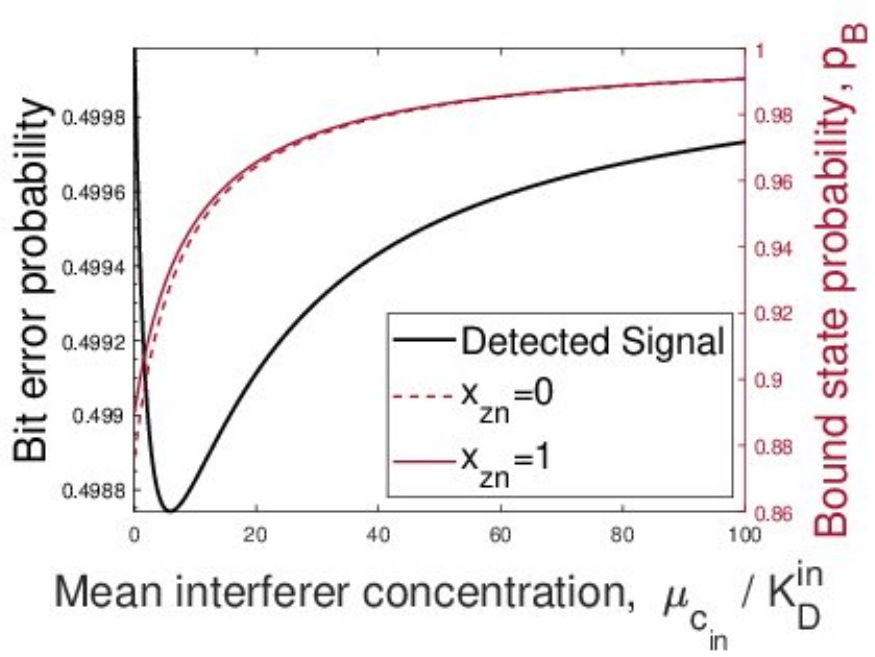}
\caption{}
\end{subfigure}
\hfill 
\begin{subfigure}{0.33\textwidth}
\centering
\includegraphics[scale=0.34]{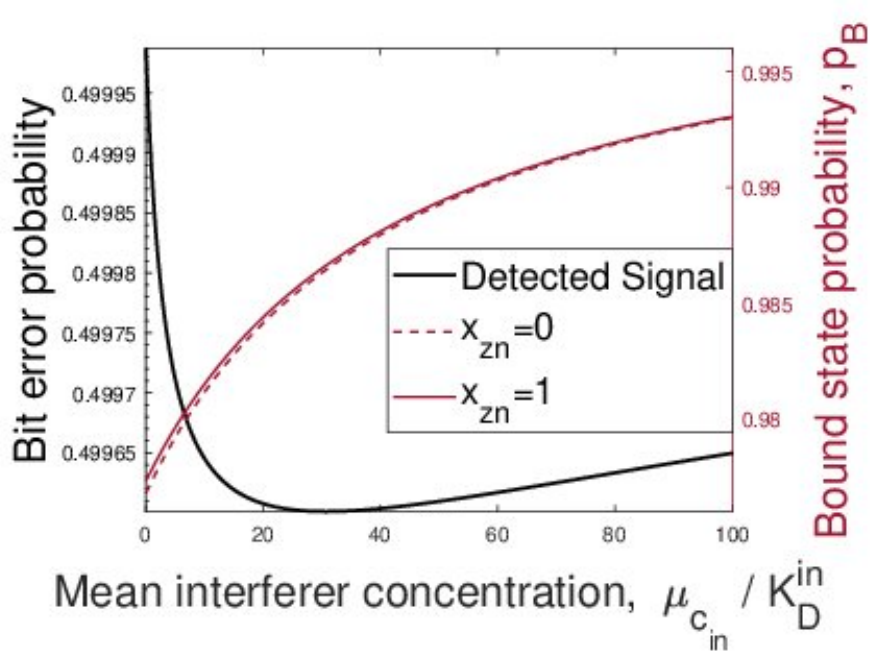} 
\caption{}
\end{subfigure}
\begin{subfigure}{0.33\textwidth}
\centering
\includegraphics[scale=0.34]{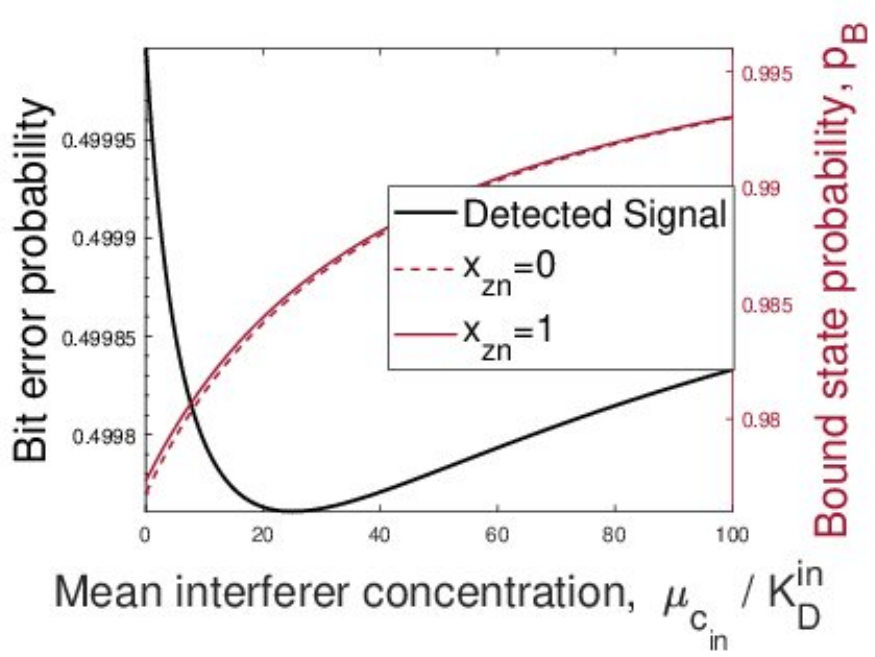} 
\caption{}
\end{subfigure}
\begin{subfigure}{0.33\textwidth}
\centering
\includegraphics[scale=0.34]{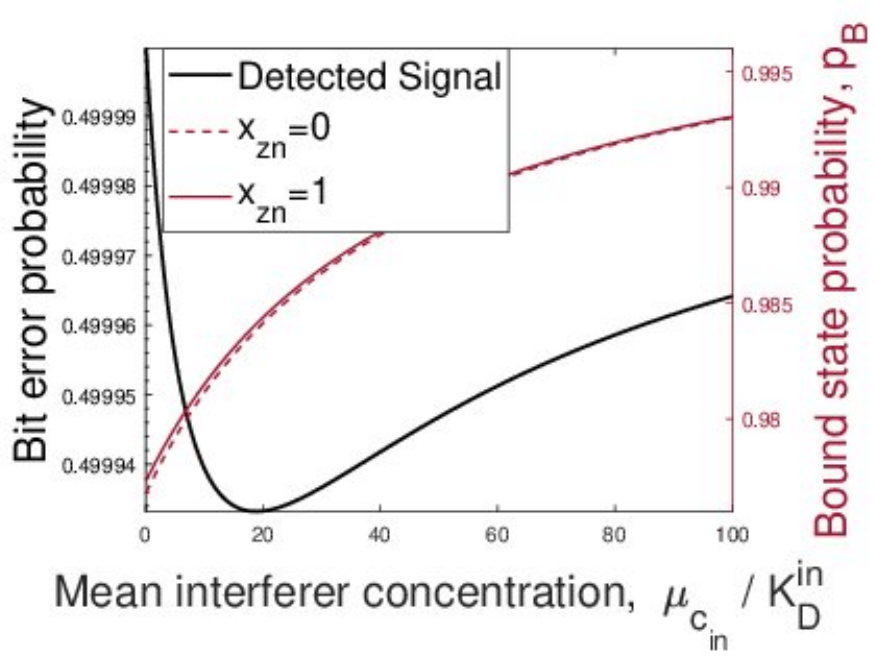}
\caption{}
\end{subfigure}
\caption{.\hspace{0.1 cm} BEP and bound state probability vs. mean interferer concentration for the non-saturation condition, where $c_{x_{zn}=0} = 4 \times Kd_{zn}$ and $c_{x_{zn}=1} = 5 \times Kd_{zn}$: in (a) with $\eta=0.2$, in (b) with $\eta=0.5$, and in (c) with $\eta=0.8$; and for the saturation condition, where $c_{x_{zn}=0} = 39 \times Kd_{zn}$ and $c_{x_{zn}=1} = 40 \times Kd_{zn}$: in (d) with $\eta=0.2$, in (e) with $\eta=0.5$, and in (f) with $\eta=0.8$.}
\label{fig:BER_Mean}   
\end{figure*}

Fig. \ref{fig:BER_Mean} shows the relationship between BEP and bound state probability as a function of mean interferer concentration, with the x-axis representing interferer concentration. The six subplots depict different affinity ratios for bit-0 (non-toxic condition) and bit-1 (toxic condition) Zn concentrations, comparing non-saturation (top row) and saturation (bottom row) conditions in the PaCoT. In non-saturation conditions, BEP is lower at low interferer concentrations, especially for lower affinity ratios, while BEP increases more sharply with rising affinity ratios. Saturation conditions show a similar trend with a more gradual increase in BEP. Overall, non-saturation conditions offer better detection but show a steeper BEP increase with rising affinity ratios than saturation.

\section{Conclusion} \label{section5}
This study presents a novel framework for mitigating heavy metal contamination using PaCoT, a device that integrates nanotechnology, biomedical engineering, and MC principles. By modeling heavy metal concentrations as a binary CSK modulated signal within the MC channel (e.g., blood vessel), we developed a method to infer toxicity conditions through precise concentration ratio estimates. The PaCoT’s reception mechanism employs predefined thresholds to distinguish between toxic and safe conditions, which are evaluated using metrics like BEP to assess detection accuracy and system reliability. Furthermore, the release mechanism of PaCoT functions as a transmitter, enabling the controlled delivery of collected toxic molecules through active components such as a single-disc viscous micropump. This design facilitates efficient discharge while carefully managing energy consumption, which is critical for extended operation in constrained environments. Performance evaluations demonstrate the system’s effectiveness based on BEP, release time, and energy consumption analyses. The findings suggest that optimizing the parameters of the PaCoT system improves both its sensitivity and operational efficiency, making it a promising solution for mitigating heavy metal exposure in human circulatory system.

\ifCLASSOPTIONcaptionsoff
  \newpage
\fi


\begin{thebibliography}{10}

\bibitem{prakash2023nano}
P.~Prakash and S.~C. S, ``Nano-phytoremediation of heavy metals from soil: a critical review,'' {\em Pollutants}, vol.~3, no.~3, pp.~360--380, 2023.

\bibitem{ekrami2022nanotechnology}
E.~Ekrami, M.~Pouresmaieli, E.~sadat Hashemiyoon, N.~Noorbakhsh, and M.~Mahmoudifard, ``Nanotechnology: a sustainable solution for heavy metals remediation,'' {\em Environmental Nanotechnology, Monitoring \& Management}, vol.~18, p.~100718, 2022.

\bibitem{baby2022nanomaterials}
R.~Baby, M.~Z. Hussein, A.~H. Abdullah, and Z.~Zainal, ``Nanomaterials for the treatment of heavy metal contaminated water,'' {\em Polymers}, vol.~14, no.~3, p.~583, 2022.

\bibitem{world2009children}
W.~H. Organization {\em et~al.}, ``Children's health and the environment. who training package for the health sector-world health organization,'' {\em http://www. who. int/ceh}, 2009.

\bibitem{soliman2022trophic}
M.~M. Soliman, T.~Hesselberg, A.~A. Mohamed, and D.~Renault, ``Trophic transfer of heavy metals along a pollution gradient in a terrestrial agro-industrial food web,'' {\em Geoderma}, vol.~413, p.~115748, 2022.

\bibitem{inobeme2023recent}
A.~Inobeme, J.~T. Mathew, C.~O. Adetunji, A.~I. Ajai, J.~Inobeme, M.~Maliki, S.~Okonkwo, M.~A. Adekoya, M.~O. Bamigboye, J.~O. Jacob, {\em et~al.}, ``Recent advances in nanotechnology for remediation of heavy metals,'' {\em Environmental monitoring and assessment}, vol.~195, no.~1, p.~111, 2023.

\bibitem{munir2021heavy}
N.~Munir, M.~Jahangeer, A.~Bouyahya, N.~El~Omari, R.~Ghchime, A.~Balahbib, S.~Aboulaghras, Z.~Mahmood, M.~Akram, S.~M. Ali~Shah, {\em et~al.}, ``Heavy metal contamination of natural foods is a serious health issue: A review,'' {\em Sustainability}, vol.~14, no.~1, p.~161, 2021.

\bibitem{gaur2021sustainable}
V.~K. Gaur, P.~Sharma, P.~Gaur, S.~Varjani, H.~H. Ngo, W.~Guo, P.~Chaturvedi, and R.~R. Singhania, ``Sustainable mitigation of heavy metals from effluents: toxicity and fate with recent technological advancements,'' {\em Bioengineered}, vol.~12, no.~1, pp.~7297--7313, 2021.

\bibitem{newmanphytoremediation}
L.~Newman, A.~A. Ansari, S.~S. Gill, M.~Naeem, and R.~Gill, ``Phytoremediation.''

\bibitem{khan2019nanoparticles}
I.~Khan, K.~Saeed, and I.~Khan, ``Nanoparticles: Properties, applications and toxicities,'' {\em Arabian journal of chemistry}, vol.~12, no.~7, pp.~908--931, 2019.

\bibitem{sanchez2011controlled}
S.~Sanchez, A.~A. Solovev, S.~Schulze, and O.~G. Schmidt, ``Controlled manipulation of multiple cells using catalytic microbots,'' {\em Chemical Communications}, vol.~47, no.~2, pp.~698--700, 2011.

\bibitem{kuscu2021internet}
M.~Kuscu and B.~D. Unluturk, ``Internet of bio-nano things: A review of applications, enabling technologies and key challenges,'' {\em arXiv preprint arXiv:2112.09249}, 2021.

\bibitem{suda2005exploratory}
T.~Suda, M.~Moore, T.~Nakano, R.~Egashira, A.~Enomoto, S.~Hiyama, and Y.~Moritani, ``Exploratory research on molecular communication between nanomachines,'' in {\em Genetic and Evolutionary Computation Conference (GECCO), Late Breaking Papers}, vol.~25, p.~29, Citeseer, 2005.

\bibitem{akyildiz2008nanonetworks}
I.~F. Akyildiz, F.~Brunetti, and C.~Bl{\'a}zquez, ``Nanonetworks: A new communication paradigm,'' {\em Computer Networks}, vol.~52, no.~12, pp.~2260--2279, 2008.

\bibitem{vilela2016graphene}
D.~Vilela, J.~Parmar, Y.~Zeng, Y.~Zhao, and S.~S{\'a}nchez, ``Graphene-based microbots for toxic heavy metal removal and recovery from water,'' {\em Nano letters}, vol.~16, no.~4, pp.~2860--2866, 2016.

\bibitem{zhang2019micro}
Y.~Zhang, K.~Yuan, and L.~Zhang, ``Micro/nanomachines: from functionalization to sensing and removal,'' {\em Advanced Materials Technologies}, vol.~4, no.~4, p.~1800636, 2019.

\bibitem{maric2018nanorobots}
T.~Maric, C.~C. Mayorga-Martinez, B.~Khezri, M.~Z.~M. Nasir, X.~Chia, and M.~Pumera, ``Nanorobots constructed from nanoclay: using nature to create self-propelled autonomous nanomachines,'' {\em Advanced Functional Materials}, vol.~28, no.~40, p.~1802762, 2018.

\bibitem{vu2018mechanisms}
T.~H. Vu and N.~Gowripalan, ``Mechanisms of heavy metal immobilisation using geopolymerisation techniques--a review,'' {\em Journal of Advanced Concrete Technology}, vol.~16, no.~3, pp.~124--135, 2018.

\bibitem{jadaa2023heavy}
W.~Jadaa and H.~Mohammed, ``Heavy metals--definition, natural and anthropogenic sources of releasing into ecosystems, toxicity, and removal methods--an overview study,'' {\em Journal of Ecological Engineering}, vol.~24, no.~6, 2023.

\bibitem{kuscu2022detection}
M.~Kuscu and O.~B. Akan, ``Detection in molecular communications with ligand receptors under molecular interference,'' {\em Digital Signal Processing}, vol.~124, p.~103186, 2022.

\bibitem{berg1977physics}
H.~C. Berg and E.~M. Purcell, ``Physics of chemoreception,'' {\em Biophysical journal}, vol.~20, no.~2, pp.~193--219, 1977.

\bibitem{ten2016fundamental}
P.~R. ten Wolde, N.~B. Becker, T.~E. Ouldridge, and A.~Mugler, ``Fundamental limits to cellular sensing,'' {\em Journal of Statistical Physics}, vol.~162, pp.~1395--1424, 2016.

\bibitem{liggett2010continuous}
T.~M. Liggett, {\em Continuous time Markov processes: an introduction}, vol.~113.
\newblock American Mathematical Soc., 2010.

\bibitem{kuscu2019channel}
M.~Kuscu and O.~B. Akan, ``Channel sensing in molecular communications with single type of ligand receptors,'' {\em IEEE Transactions on Communications}, vol.~67, no.~10, pp.~6868--6884, 2019.

\bibitem{bian2016111}
X.~Bian, C.~Kim, and G.~E. Karniadakis, ``111 years of brownian motion,'' {\em Soft Matter}, vol.~12, no.~30, pp.~6331--6346, 2016.

\bibitem{blanchard2006miniature}
D.~Blanchard, P.~Ligrani, and B.~Gale, ``Miniature single-disk viscous pump (single-dvp), performance characterization,'' 2006.

\bibitem{al2007investigation}
A.~Al-Halhouli, M.~Kilani, A.~Al-Salaymeh, and S.~B{\"u}ttgenbach, ``Investigation of the influence of design parameters on the flow performance of single and double disk viscous micropumps,'' {\em Microsystem technologies}, vol.~13, pp.~677--687, 2007.

\bibitem{alparametric}
A.~Al-Halhouli, M.~Kilani, M.~Amayreh, A.~Al-Salaymeh, and S.~B{\"u}ttgenbach, ``Parametric study of single disk viscous micropump,''

\bibitem{al2009recent}
A.~T. Al-Halhouli, ``Recent advances in on-disk viscous micropumps,'' {\em Journal of microelectronics and electronic packaging}, vol.~6, no.~4, pp.~240--249, 2009.

\bibitem{korkola2024structural}
N.~C. Korkola and M.~J. Stillman, ``Structural motifs in the early metallation steps of zn (ii) and cd (ii) binding to apo-metallothionein 1a,'' {\em Journal of Inorganic Biochemistry}, vol.~251, p.~112429, 2024.

\bibitem{blanchard2005single}
D.~Blanchard, P.~Ligrani, and B.~Gale, ``Single-disk and double-disk viscous micropumps,'' {\em Sensors and Actuators A: Physical}, vol.~122, no.~1, pp.~149--158, 2005.

\bibitem{balfourier2022importance}
A.~Balfourier, A.-P. Marty, and F.~Gazeau, ``Importance of metal biotransformation in cell response to metallic nanoparticles: A transcriptomic meta-analysis study,'' {\em ACS Nanoscience Au}, vol.~3, no.~1, pp.~46--57, 2022.

\end{thebibliography}
\end{document}